# ATP-driven separation of liquid phase condensates in bacteria

B. Guilhas[1], J.C. Walter[2], J. Rech[3], G. David[2], N.-O. Walliser[2], J. Palmeri[2], C., Mathieu-Demaziere[3], A. Parmeggiani[2,4], J.Y. Bouet[3], A. Le Gall[1#], M. Nollmann[1#]

[1] Centre de Biochimie Structurale, CNRS UMR 5048, INSERM U1054, Université de Montpellier, 60 rue de Navacelles, 34090, Montpellier, France

[2] Laboratoire Charles Coulomb (L2C), Univ. Montpellier, CNRS, Montpellier, France.

[3] LMGM, CBI, CNRS, Univ. Toulouse, UPS, Toulouse, France.

[4] LPHI, CNRS, Univ. Montpellier, Montpellier, France.

# Corresponding authors

## Abstract

Liquid-liquid phase separated (LLPS) states are key to compartmentalise components in the absence of membranes, however it is unclear whether LLPS condensates are actively and specifically organized in the sub-cellular space and by which mechanisms. Here, we address this question by focusing on the ParAB*S* DNA segregation system, composed of a centromeric-like sequence (*parS*), a DNA-binding protein (ParB) and a motor (ParA). We show that *parS*-ParB associate to form nanometer-sized, round condensates. ParB molecules diffuse rapidly within the nucleoid volume, but display confined motions when trapped inside ParB condensates. Single ParB molecules are able to rapidly diffuse between different condensates, and nucleation is strongly favoured by *parS*. Notably, the ParA motor is required to prevent the fusion of ParB condensates. These results describe a novel active mechanism that splits, segregates and localises non-canonical LLPS condensates in the sub-cellular space.

## Keywords

## Introduction

In the past, bacteria were often viewed as homogeneous systems lacking the complex sub-cellular spatial organization patterns observed in eukaryotes. The advent of powerful labeling and imaging methods has enabled the discovery of a plethora of mechanisms used by bacteria to specifically and precisely localize components, in space and time, within their sub-cellular volume (Surovtsev and Jacobs-Wagner 2018; Shapiro, McAdams, and Losick 2009). These mechanisms include pole-to-pole oscillatory systems to define the site of cell division (e.g. MinCDE), dynamic, ATP-driven polymerization to drive cell division and cell growth (e.g FtsZ, MreB), recognition of cell curvature to localize chemotaxis complexes (e.g. DivIVA, (Ramamurthi and Losick 2009)), ATP-powered translocation of membrane-bound machines to power cell motility (e.g. Agl-Glt (Faure et al. 2016)), or nucleoid-bound oscillatory systems to localize chromosomal loci (e.g. ParABS, (Le Gall et al. 2016)). More recently, it became apparent that many cellular components (e.g. ribosomes, RNA polymerases, P-granules) (Sanamrad et al. 2014; van Gijtenbeek et al. 2016; Moffitt et al. 2016; Racki et al. 2017) display specific sub-cellular localization patterns, leading to the spatial segregation of biochemical reactions (e.g. translation, transcription, or polyP biosynthesis). Most notably, bacteria are able to achieve this precise sub-cellular compartmentalization without resorting to membrane-enclosed organelles.

Recently, important progress has been made in understanding the formation of membrane-less organelles in eukaryotic cells (Hyman, Weber, and Jülicher 2014). A combination of *in vitro* and *in vivo* experiments demonstrated that such compartments are formed by liquid-liquid phase separation (LLPS) (Hyman, Weber, and Jülicher 2014), a thermodynamic process by which attractive molecular interactions counterbalance entropy-driven effects. This phenomenon promotes the self-organisation of a condensed phase, in which molecules are close enough from each other to experience their mutual attraction, interfaced with a dilute phase. This mechanism provides advantages such as rapid assembly/disassembly, absence of a breakable membrane, and serves fundamental biological processes such as regulation of biochemical reactions, sequestration of toxic factors, or organization of hubs (Shin and Brangwynne 2017).

The first evidence that eukaryotic cells use LLPS came from examination of P granules in *C. elegans* (Brangwynne et al. 2009). In this study, Brangwynne et al. observed different key signatures of liquid droplets: P granules formed spherical bodies that could fuse together, drip and wet, and displayed a dynamic internal organisation. Since then, many other processes in eukaryotes were shown to display LLPS properties (Banani et al. 2017). A few recent examples show that this phenomenon also exists in bacteria, for instance ribonucleoprotein granules (Al-Husini et al. 2018), the cell division protein FtsZ (Monterroso et al. 2019), and carboxysomes (Wang et al. 2019; MacCready, Basalla, and Vecchiarelli 2020). Thus, LLPS seems to be a universal mechanism to compartmentalise components in the absence of membranes. However, it is unclear whether LLPS condensates are actively and specifically organized in the sub-cellular space and by which mechanisms.

We addressed this problem by investigating how specific DNA sequences are organized within the sub-cellular volume in space and time. We focused on the ParABS partition system, responsible for chromosome and plasmid segregation in bacteria and archaea (Bouet et al. 2014; Toro and Shapiro 2010; Baxter and Funnell 2014; Schumacher et al. 2015). This system is composed of three components: (1) a centromeric sequence (*parS*); (2) a dimeric DNA binding protein (ParB) that binds *parS;* and (3) a Walker A ATPase (ParA). We have previously shown that ParB is very abundant (>850 dimers per cell) (Bouet et al.

2005) and localizes to a large extent (~90% of ParB molecules) to regions containing *parS* (Sanchez et al. 2015) to form a tight nucleoprotein complex (partition complex). In addition to specific interactions with *parS,* other ingredients were shown to be involved in the assembly of the partition complex: weak ParB-ParB dimer interactions mediated by its disordered, low complexity region, as well as non-specific interactions with the surrounding DNA (Debaugny et al. 2018). Different models were proposed for the assembly of partition complexes. The 'spreading' model postulates that ParB dimers load at *parS* sites and then spread outwards over large genomic distances (Murray, Ferreira, and Errington 2006; Rodionov, Lobocka, and Yarmolinsky 1999). In the 'spreading and bridging' and 'caging' models, additional long-range interactions between ParB dimers and DNA help the partition complex assemble and coalesce into compact three dimensional assemblies (Broedersz et al. 2014; Sanchez et al. 2015; Graham et al. 2014).

Here, we used super-resolution microscopy and single-particle analysis to investigate the physical mechanisms involved in the formation of bacterial partition complexes. We show that partition complexes are small (<50nm), spherical objects. Single, isolated ParB molecules diffuse rapidly within the nucleoid volume, but display confined motions when trapped inside partition complexes. These results suggest a partition of ParB into two phases: a gas-like phase, and a dense, liquid-like condensate phase that shows nanometer-sized, spherical droplets. Next, we show that the nucleation of ParB condensates is strongly favoured by the presence of the centromeric sequence *parS* and that separation and proper sub-cellular localization of ParB condensates require the ParA motor. Critically, different ParB condensates collapse into a single condensate upon degradation of ParA or by mutating ParA's ATPase nucleotide-binding site. These results describe a novel active mechanism that splits, segregates and localises LLPS condensates in the sub-cellular space.

# Results

## ParB assembles into spherical nano-condensates

Previous studies revealed that the partition complex is made of ~250 dimers of ParB (Adachi, Hori, and Hiraga 2006; Bouet et al. 2005) and ~10kb of *parS*-proximal DNA (Rodionov, Lobocka, and Yarmolinsky 1999). This complex is held together by specific, high-affinity interactions between ParB and *parS*, and by low-affinity interactions between ParB dimers that are essential for the distribution of ParB around *parS* sites (Figure 1A) (Sanchez et al. 2015; Debaugny et al. 2018). However, technological limitations have thwarted the investigation of the physical mechanisms involved in the formation of partition complexes.

We addressed these limitations by first investigating the shape and size of the F-plasmid partition complex, reasoning that the former should inform us on the role of the mechanisms involved in the maintenance of the complex cohesion while the latter would enable an estimation of protein concentration within the partition complex. To this aim, we combined Photo-Activated Localisation Microscopy (PALM) (Marbouty et al. 2015; Fiche et al. 2013) with single-particle reconstruction (Salas et al. 2017), and used a previously validated functional ParB-mEos2 fusion strain (Sanchez et al. 2015). In addition, we implemented an efficient and well-controlled fixation protocol (Figure S1A and Methods) to limit the influence of partition complex dynamics and reach a localization precision of $14 \pm 6\ nm$ (Figures S1B-C). Most single ParB particles localized to the partition complex, as we previously observed by live PALM (Sanchez et al. 2015) (Figure 1B). Next, we clustered localizations pertaining to each partition complex using a density-based clusterization algorithm (Cattoni et al. 2017; Levet et al. 2015). Partition complexes were positioned near the quarter cell positions (Figure S1A), as reported previously (Le Gall et al. 2016). Partition complex shapes displayed heterogeneity, reflecting the different three dimensional projections and conformational states of the complex (Figure 1B-C). Thus, we used single-particle analysis to classify them into eight major class averages that contained most partition complexes (Figure 1C, S1D-E) and applied several quantitative methods to assess particle shape in each class (Figures S1G-I). Direct comparison with simulated spherical particles revealed that experimental particles within each class display highly symmetric shapes with very little roughness. From the first eight class averages, we generated 3D reconstructions reaching nanometer-scale isotropic resolution (Figure 1D) (Salas et al. 2017).

Next, we estimated the size of the partition complex by calculating the mean full width at half maximum of the reconstructions obtained from each single class average (average of all classes = 43 $\pm$ 7 nm) (Figure S1F). Thus, from the size of the partition complex and the average number of ParB in each partition complex (~250 ParB dimers) (Adachi, Hori, and Hiraga 2006; Bouet et al. 2005), we can estimate a local ParB dimer concentration of the order of ~10 mM (see Methods for details). Remarkably, this extremely high concentration is comparable to that of the total protein concentration in the bacterial cytoplasm (~10 mM, for proteins of the same size as ParB) (Elowitz et al. 1999), suggesting that ParB dimers represent most of the total protein volume within a partition complex. ParB dimers interact together with sub-$\mu$M affinities (0.3-1 $\mu$M) (Taylor et al. 2015; Sanchez et al. 2015), thus we expect virtually every ParB dimer within a partition complex to be interacting with another ParB dimer. Overall, these data show that ParB assembles nanometer-sized, high-concentration, membrane-free spherical compartments.

**ParB exists in an equilibrium between liquid- and gas-like phases**

Next, we investigated whether single ParB dimers were able to escape partition complexes by using single-particle tracking PALM (sptPALM) (Le Gall et al. 2016; Sanchez et al. 2015; Manley et al. 2008; Niu and Yu 2008). We observed two distinct dynamic behaviors reflected by low-mobility (or confined) and high-mobility trajectories with clearly different apparent diffusional properties (Figures 2A-B). The low-mobility species is consistent with immobile ParB dimers bound to DNA specifically, whereas the high-mobility fraction is consistent with ParB dimers interacting non-specifically with chromosomal DNA (see discussion in Figure S2).

A recent study explored the dynamic behaviour of LacI in conditions where the number of LacI molecules was in vast excess with respect to the number of specific Lac repressor sites, mirroring the experimental conditions found in the ParB*S* system (Garza de Leon et al. 2017). In this case, the authors observed that ~80% of trajectories were highly mobile, corresponding to repressors interacting non-specifically with DNA, with only ~20% of tracks displaying low mobility and corresponding to repressors bound to their operators. In stark contrast, for the ParAB*S* system, most ParB particles displayed a low-mobility (~95%), and only a small proportion of particles were highly mobile (<5%) (Figure 2B).

Low-mobility trajectories localized near the quarter cell positions, similarly to ParB complexes (Figure S2A-B). To determine whether these trajectories clustered in space, we calculated the pair correlation between low-mobility trajectories (see Methods). For a homogeneous distribution we would expect a flat line (see dashed curve in Figure 2C). In contrast, we observed a sharp peak at short distances, indicating that low-mobility ParB particles are spatially confined within partition complexes (Figure 2C, blue curve). High-mobility trajectories, instead, occupied the whole nucleoid space (Figure S2A). Thus, from the percentage of high-mobility trajectories detected (~5% of all trajectories), and the estimated number of ParB dimers per cell (~800), one can estimate that only ~20 ParB dimers are typically moving between partition complexes. High-mobility ParB dimers move on the nucleoid volume, thus we estimate that this species exists at a concentration of ~20 nM, five orders of magnitude smaller than the ParB concentration within partition complexes (~10 mM). Therefore, these results suggest that ParB molecules exist in two well-defined phases: (1) a highly condensed, liquid-like state (ParB condensate) containing a high concentration of low-mobility ParB dimers; and (2) a diluted gas-like phase where single, high-mobility ParB dimers diffuse on the nucleoid by weak non-specific DNA interactions (Sanchez et al. 2013; Taylor et al. 2015; Fisher et al. 2017).

**ParB diffuses across phase boundaries**

This interpretation suggests that single ParB molecules should be able to diffuse between two different partition complexes within the same cell. To test this, we photobleached a single partition complex and monitored the total fluorescence signal in both the bleached and unbleached complex (Figure 2D). Strikingly, we observed a clear increase in the fluorescence signal of the bleached complex over time with a concomitant decrease of the fluorescence signal of the unbleached complex (Figures 2D-E). This behaviour can only be explained by the escape of fluorescent proteins from the unbleached complex into the gas-phase, and the entry of proteins from the gas-phase into the bleached complex. This produces a net flux between unbleached and bleached complexes and vice-versa. The symmetry between both curves suggests that both fluxes have similar magnitudes (Figure 2E, S2C). By modeling the diffusion process, we could fit the experimental curves to a model and estimated the residence time of a ParB molecule in a ParB condensate to be ~ 100 s (Figures S2D-F), while the typical time in the gas-like phase was ~23 s. Note that these typical times provide a good estimate (~90%) of the number of ParB dimers confined within partition complexes (see Figure S2).

Eventually, the system reached equilibrium (after ~4 min). At this point, the fluorescence intensities of bleached and unbleached complexes became equal, indicating that ParB molecules equipartitioned between the two condensates. We note that equipartition between ParB condensates would only be possible if single ParB dimers were diffusing across the liquid-like and gas-like phase boundaries, therefore equilibrating chemical potentials between these two phases (Hyman, Weber, and Jülicher 2014). In equilibrium, small condensates in a canonical LLPS system are expected to fuse into a single, large condensate, for instance by Ostwald ripening (Hyman et al., 2014). Thus, the observation that ParB condensates are in mutual stationary equilibrium at times smaller than the cell cycle suggests non-canonical mechanisms to suppress droplet fusion.

**ParB condensates are able to merge**

Biomolecular condensates can exhibit different internal architectures. For example, the Balbiani body is a solid-like structure maintained by a network of amyloid-type fibers (Shin and Brangwynne 2017), while liquid-like behavior of P granules denotes an internal disordered organization. Thus, the ability of two compartments to fuse helps distinguish between these two different internal structures. To study whether ParB condensates were able to fuse over time, we performed live time-lapse fluorescence microscopy experiments. We observed that independent partition complexes were able to fuse into a single partition complex (Figure 2F). Notably, we observed that fused complexes move in concert during several seconds (Figure 2F). These fusion events were rapid (~5±3 s) and reversible (14 out of 14 events). To estimate whether the number of ParB molecules was conserved upon fusion, we integrated the fluorescence intensity signal of each ParB condensate prior and after fusion. As expected, the integrated intensity increased upon fusion (Figure S2G) but only by 1.5±0.1 fold, instead of the two-fold increase expected for canonical LLPS systems. Conversely, splitting of ParB condensates decreased the integrated intensity by 0.7±0.1 fold (Figure S2H). We note that this ability of ParB complexes to fuse is not unique to the F-plasmid system, and was also observed for P1 plasmids which also segregate using a ParABS system (Sengupta et al. 2010).

**_parS,_ and low- and high-affinity ParB interactions are sufficient for phase separation**

Next, we performed a thermodynamic analysis and Monte-Carlo simulations to find the minimal set of assumptions that would be necessary to reproduce an equilibrium between liquid- and gas-like phases. For this, we considered a minimalistic gas-liquid lattice model that shows in mean field theory a generic LLPS diagram (Figure S3A). In our experimental conditions, the percentage loss of plasmid in a Δ ParAB mutant strain can be used to estimate on average ~1 plasmid/ParB condensate. Thus, we simulated 300 ParB particles on a three-dimensional 2x0.5x0.5 $\mu m^3$ lattice with a spacing of 5 nm, and represented the F-plasmid by a single, static, high affinity binding site within this lattice containing a repeat of 10 _parS_ sequences. ParB interacted with high affinity with the _parS_ cluster (hereafter _parS10_), with low-affinity with all other sites in the lattice, and with low-affinity with itself. Affinities and concentrations were based on experimentally-determined coefficients (Figure S3). The system was left to evolve under different initial conditions (pure gas-like or liquid-like states) and in the presence or absence of _parS10_. The system remained in a gas-like phase in the absence of _parS10_ (Figure 3A, green curve) or when ParB-ParB interactions were too weak (Figure S3D-E). In contrast, in the presence of *parS10* and of physiological ParB-ParB interactions the system sequentially evolved towards states involving *parS* binding, nucleation, and stable co-existence of liquid- and gas-like phases (Figure 3A, blue curve). The system evolved towards the same endpoint when the initial condition was a pure liquid-like state (Figure 3A, gray curve). At this endpoint, 80% of the molecules were in a liquid-phase and the remainder in a gas-like state, comparable to the proportion of clustered, low-mobility trajectories observed experimentally (95%, Figure 2B). Thus, this minimalist model reproduces the basic experimental phenomenology and shows that the only elements required to reach a stable coexistence between liquid and gas-like ParB phases are: low-affinity interactions between ParB dimers, and high-affinity interactions between ParB and the nucleation sequence *parS*. Critically, ParB was unable to form a stable liquid-gas coexistence phase in the absence of *parS* within physiological timescales (Figure 3A, green curve), whilst the nucleation of ParB condensates was accelerated by an increasing number of *parS* sequences (Figure S3F). Thus, our simulations suggest that efficient nucleation of ParB into condensates requires *parS*.

We experimentally tested this prediction by performing sptPALM experiments in a strain lacking *parS*. In contrast to our previous results, we frequently observed high-mobility trajectories (Figure 3B-C). In fact, the proportion of this population increased from ~5% in the wild-type to ~50% in the absence of *parS*. The persistence of low-mobility trajectories suggests that a fraction of ParB dimers can remain bound to other chromosomal sequences mimicking *parS* sites. To test this hypothesis, we first searched the *E.coli* chromosome for sequences displaying homology to the *parS* motif (TGGGACCACGGTCCCA) using FIMO and found tens of sequences displaying only two or three mutations in the consensus *parS* sequence (Figure S3G). Second, we observed using Monte-Carlo simulations that single, parS-like chromosomal sequences can be partially occupied in the absence of parS (Figure S3H). Finally, we found using sptPALM that a ParB mutant with reduced parS-binding affinity exhibits a further reduction in the low-mobility population with respect to the strain lacking parS (from 50% to 30%, Figure S3I). Overall, these results indicate that weakening ParB-parS interactions leads to a large shift of ParB dimers towards the gas phase. However, it was not clear whether the remaining low-mobility ParB dimers still formed condensed phases.

To address this question, we performed pair-correlation analysis of low-mobility trajectories (Figure 3D). Clearly, low-mobility trajectories displayed a homogeneous distribution and failed to cluster in both mutant strains, in contrast with the clustering behaviour of low-mobility trajectories in the wild-type (Figure 3D). All in all, these observations show that formation of ParB condensates requires ParB-parS interactions, and are consistent with previous studies *(Erdmann, Petroff, and Funnell 1999; Sanchez et al. 2013)(Sanchez et al. 2013)*

A second conclusion of our Monte-Carlo model is that ParB dimer interactions are also required for the formation of a condensed phase. To test this prediction experimentally, we used a strain carrying a mutation in the arginine-rich motif of ParB that partially reduces ParB dimer interactions (ParB-3R* mutant). In this mutant, formation of partition complexes is considerably impaired (Debaugny et al. 2018). By sptPALM, we observed that low-mobility trajectories considerably decrease their ability to cluster (Figure 3D), consistent with previous studies (Debaugny et al., 2018; Song et al., 2017), and in support of our Monte Carlo simulations (Figure S3D-E). Taken together, these data indicate that high-affinity ParB-parS interactions, and low-affinity ParB dimer interactions are needed for the assembly of stable liquid-like ParB condensates. Weak interactions between ParB dimers and nsDNA were also shown to be necessary for partition complex assembly (Sanchez et al. 2013; Taylor et al. 2015; Fisher et al. 2017).

## Motor proteins drive ParB condensates out-of-equilibrium

At equilibrium, a passive system undergoing phase separation displays a single condensed phase: if initially multiple condensed phases are found, the lowest energy state comprises a single liquid-separated phase that can be reached either by fusion or through Ostwald ripening (Zwicker et al. 2014). Partition systems have specifically evolved to ensure faithful DNA segregation to daughter cells, which would be impeded by fusion into single, stable partition complexes. Our previous time-lapse fluorescence microscopy experiments showed that single partition complexes can merge at short time-scales (tens of seconds, Figure 2F). At longer time-scales (tens of minutes), however, ParB condensates are not only kept apart from each other but are also actively segregated concomitantly with the growth of the host cell (Figure 4A). Previous reports have shown that the ParA motor protein is necessary to ensure faithful separation and segregation of partition complexes (Le Gall et al. 2016; Erdmann, Petroff, and Funnell 1999). This function requires both ParA's ATP hydrolysis activity and its stimulation by ParB (Le Gall et al. 2016; Ah-Seng et al. 2013), and would ensure the segregation of ParB condensates. However, it is not clear whether ParA may play a role at preventing fusion of ParB condensates.

If ParA was the factor preventing fusion of ParB condensates, then its slow removal should lead to the fusion of all ParB condensates in a cell. To test this hypothesis, we generated a strain where ParA can be selectively degraded using a degron system (McGinness, Baker, and Sauer 2006) (Figure 4B). In this strain, we observed a large reduction in ParA levels after 30 min of degron induction (Figure 4C). In the absence of ParA degradation, this strain displayed wildtype ParB condensate separation dynamics (Figure 4D, white traces). Strikingly, upon degradation of ParA, we observed the fusion of ParB condensates (Figure 4E, white traces). Over time, the average number of ParB condensates per cell collapsed to 0.6 ± 0.7 ParB condensates/cell in ParA-degraded cells, but remained at 2.9 ± 1.0 ParB condensates/cell in cells where ParA degradation was not induced (Figure 4F, S4A). To test whether this reduction was due to the fusion of partition complexes, we quantified the mean intensity of ParB condensates before (0h) and 3h after ParA degradation (Figure S4B). The marked increase in the mean intensity of ParB condensates over time (from 2.9±0.9 at 0h to 4.9± 2.3 at 3h) further suggests that the observed decrease in the number of ParB condensates was due to condensate fusion.

Next, we wondered whether ParA degradation would have other consequences in ParB condensate dynamics. For this, we quantified the sub-cellular positioning of ParB condensates before (0h) and 3h after ParA degradation by calculating the localization density histogram of ParB condensates along the longitudinal axis of the cell. These histograms show that ParA degradation changes the average sub-cellular positioning of ParB condensates (Figure 4F), in good agreement with previous findings using ParA deletion strains (Le Gall et al. 2016).

While these results show that ParA is necessary to maintain ParB condensates apart, they do not provide direct evidence for the need of ParA's ATPase activity in this process. For this, we quantified the number of partition complexes in cells where we directly mutated ParA's ATPase nucleotide-binding site (ParA$^{K120Q}$, a mutant impaired in ATP hydrolysis stimulated by ParB). The mean number of partition complexes was 3.0±1.1 in wild-type cells, and decreased to 1.8±0.9 in the ParA ATPase mutant strain (Figure 4G), consistent with ParA's ATPase activity being implicated in the separation of ParB condensates.

## Discussion

Here, we provide evidence in support of a new class of droplet-like-forming system with unique properties that ensure the stable coexistence and regulated inheritance of separate liquid-condensates. Interestingly, the three components of the ParABS partition system and an exquisite regulation of their multivalent interactions are required to observe this complex behavior: (1) High-affinity interactions between ParB and the *parS* centromeric sequence are necessary for the nucleation of the ParB condensate; (2) Weak interactions of ParB with chromosomal DNA and with itself are necessary to produce phase separation; (3) The ParA ATPase is required to counter condensate fusion and to properly position condensates in the cell. Interestingly, this system displays several non-canonical properties discussed below.

*In vivo*, ParB binds *parS* and an extended genomic region of ~10kb around *parS* (Murray, Ferreira, and Errington 2006; Rodionov, Lobocka, and Yarmolinsky 1999; Sanchez et al. 2015). Interestingly, the size of this extended genomic region is independent of the intracellular ParB concentration (Debaugny et al. 2018), suggesting that an increase in the total number of ParB molecules per cell either leads to an increase in the concentration of ParB within droplets, or that an unknown mechanism controls the size of ParB droplets. Here, we found that the intensity of ParB condensates increases by 1.5±0.1 fold upon fusion, instead of doubling as expected for a conservative process. Consistently, the intensity of ParB condensates did not decrease by two-fold but by 0.7±0.1 upon condensate splitting. All in all, these results suggest that droplet fusion/splitting may be semi-conservative, and point to a more complex mechanism regulating ParB number and/or droplet size.

Proteins in canonical LLPS systems assemble in compact droplets, such as oil molecules in an oil-in-water emulsion. In this case, an increase in the number of 'oil' molecules leads to an increase in droplet size because the concentration of oil molecules in droplets is at its maximum (i.e the droplet is compact). There are two key differences in the ParAB*S* system that mean that a doubling in ParB number does not necessarily lead to a doubling in droplet volume. First, most ParB proteins in ParB*S* condensates are bound to the DNA lattice formed by *parS* and the surrounding chromosome/plasmid DNA. Thus, ParB mainly lives on a polymer and therefore an increase in ParB number can compact the DNA in the droplet leading to only slight increases in condensate size. Second, the concentration of ParB in ParB condensates is ~10 mM. This concentration is not that of a compact aggregate in a watery solvent. Therefore, increases in ParB numbers can lead to increases in the concentration of ParB in droplets that do not lead to noticeable changes in droplet sizes. Finally, we cannot discard mechanisms controlling the size of ParB condensates (Söding et al. 2019; Garcia-Jove Navarro et al. 2019; Zwicker et al. 2014) that could decouple increases in droplet volume from increases in the number of ParB molecules.

In passive, canonical phase-separation systems, separate liquid droplets grow by taking up material from a supersaturated environment, by Ostwald ripening, or by fusion of droplets. Over time, these processes lead to a decrease in the number of droplets and an increase in their size. This reflects the behaviour we observed for ParB condensates upon depletion of ParA or when its ATPase activity was reduced. Recently, theoretical models have predicted mechanisms to enable the stable coexistence of multiple liquid phases (Zwicker, Hyman, and Jülicher 2015). These mechanisms require constituents of the phase separating liquids to be converted into each other by nonequilibrium chemical reactions (Zwicker, Hyman, and Jülicher 2015; Zwicker et al. 2016). Our result that the ATPase activity of ParA is required to maintain ParB condensates well separated indicates that motor proteins can be involved in the active control of droplet number and in their sub-cellular localization. Interestingly, similar mechanisms –yet to be discovered– could control the number, size, and sub-cellular localization of other droplet-forming systems such as P-granules (Brangwynne 2011), stress granules (Jain et al. 2016), or heterochromatin domains (Strom et al. 2017).

## Acknowledgments

This project has received funding from the European Research Council (ERC) under the European Union's Horizon 2020 Research and Innovation Program (Grant ID 724429) (M.N.), the Agence Nationale pour la Recherche grant HiResBacs (ANR-15-CE11-0023) (M.N.), grant IBM (ANR-14-CE09-0025-01) (M.N., J-Y.B, A.P.), the ECOS-Sud program (A16B01), the Program 'Investissements d'Avenir' for the Labex NUMEV (2011-LABX-076) integrated into the I-Site MUSE (AAP 2013-2-005, 2015-2-055, 2016-1-024) (G.D., J-C.W., N-O.W., J.P., A.P.) and the CNRS INPHYNITI program (J-C.W., J-Y.B). We acknowledge the France-BioImaging infrastructure supported by the French National Research Agency (grant ID ANR-10-INBS-04, ''Investments for the Future''). B.G. and G.D. were supported by a PhD fellowship from the French Ministère de l'Enseignement Supérieur, de la Recherche et de l'Innovation.

## Data availability statement

The data that support the findings of this study are available from the corresponding author upon reasonable request.

## Code availability

Most analysis tools and methods used in this study are in the open domain (see Methods). Code developed specifically for this manuscript can be obtained from the corresponding author upon request.

# Methods

### Sample preparation

*sptPALM, PALM, FRAP and time-lapse microscopy (fast dynamics)*

Microscopy coverslips and glass slides were rinsed with acetone and sonicated in a 1M KOH solution for 20 minutes. Next, they were dried over an open flame to eliminate any remaining fluorescent contamination. A frame of double-sided adhesive tape was placed on a glass slide and a ~5mm channel was extruded from its center as previously described (Le Gall, Cattoni, and Nollmann 2017). Briefly, 20 μl of 2 % melted agarose (diluted in M9 media, melted at 80°C; for the imaging of DLT3469 strain: 1 % melted agarose diluted in M9 + 0.2% arabinose) were spread on the center of the glass slide and covered with a second glass slide to ensure a flat agarose surface. The sandwich slides were kept on a horizontal position for 5 min under external pressure at room temperature (RT) to allow the agarose to solidify. The top slide was then carefully removed when bacteria were ready to be deposited in the agar pad (see below). Cells were harvested during the exponential phase (optical density at 600 nm: ~0.3). For PALM experiments, cells were fixed in a 2% PFA solution for 15 min at room temperature and 30 min at 4°C (for the detailed procedure, refer to (Visser, Joshi, and Bates 2017)). A bacterial aliquot was spun down in a bench centrifuge at RT at 3000g for 3 minutes. The supernatant was then discarded and the pellet re-suspended in 10 μl of minimal media. 1.5 μl of 1/10 fiducial beads (TetraSpeck TM, Life Technologies) were added.

1.7 µl of the resulting bacterial solution were pipetted onto the agar. After deposition of bacteria, the second face of the double-sided tape was removed and the pad was then sealed with a clean coverslip.

### *Time-lapse microscopy (slow dynamics)*

Long-term ParB-GFP droplet dynamics (Figure 4) were monitored by following the movement of ParB droplets over several cell cycle times. For this, cells were first harvested during the exponential phase (optical density at 600 nm: ~0.3), then diluted to 1:300 to adjust the bacterial density on the field of view, and finally deposed in an ONIX CellAsic microfluidic chamber (B04A plates; Merck-Millipore). Bacteria were immobilized and fresh culture medium (supplemented with 0.2% arabinose) was provided (10.3 kPa) throughout the acquisition.

### **Fluorescence microscopy**

### *PALM and sptPALM*

PALM and sptPALM imaging of ParB-mEos2 were performed on a home-built experimental setup based on a Zeiss Axiovert 200 by continuously acquiring 7000 (20000 for sptPALM) images at 50ms exposure time  (8ms for sptPALM) with a 561 nm readout laser (Sapphire 561LP, 100mW, Coherent) and continuous illumination with a 405 nm laser for photoactivation (OBIS 40550, Coherent). Data were acquired using custom-made code written in Labview. The readout laser intensity used was 1.4 kW/cm² at the sample plane. The intensity of the 405nm laser was modified during the course of the experiment to maintain the density of activated fluorophores constant while ensuring single molecule imaging conditions. PALM acquisitions were carried out until all mEos2 proteins were photoactivated and bleached. Fluorescent beads (TetraSpeck TM, Life Technologies) were used as fiducial marks to correct for sample drift by post-processing analysis. For cell segmentation, a bright field image was taken on each field of view.

*Time-lapse microscopy*

Long-term time-lapse microscopy to monitor of ParB-GFP droplet dynamics was performed on the same optical setup by continuously acquiring images at 50ms exposure time with a 561 nm readout laser (Sapphire 561LP, 100mW, Coherent) until complete photobleaching.

## FRAP

FRAP experiments on ParB-GFP droplets were conducted on a ZEISS LSM 800 by acquiring 318x318x5 (XYZ) images, every 10 seconds, with 106x106x230 $nm^3$ voxels exposed during 1.08 $\mu$s. Data were acquired using Zen Black (Zeiss acquisition suite). The photo-bleaching of ParB-GFP droplets was performed in circular regions with a diameter of 1.2 $\mu$m covering single droplets, in cells with exactly two droplets. To be able to detect them and estimate their intensity afterwards, droplets were only partially photo-bleached (50% approximately). To estimate photo-bleaching during fluorescence recovery, fluorescent intensity was also monitored in bacteria in which ParB-GFP droplets were not pre-bleached. The resulting z-stack images were then projected and analyzed.

*ParA degradation experiments*

ParA degron experiments (DLT3469 strain) were conducted on an Eclipse TI-E/B wide field epifluorescence microscope with a phase contrast objective. To quantify the number of ParB-GFP droplets, samples were prepared at different time points ($t_{induction}$, $t_{induction}$+ 1h, $t_{induction}$ + 2h, $t_{induction}$+ 3h, and one night after induction of degradation) by harvesting cells from the solution, and snapshot images were acquired with an exposure time of 0.6 s.

Images were then analysed using the MATLAB-based open-source software MicrobeTracker, and the SpotFinderZ tool (Sliusarenko et al., 2011). To monitor the dynamics of ParB-GFP droplets, images were acquired every 42 seconds.

**Data analysis**

*PALM*

Localization of single molecules for PALM experiments was performed using 3D-DAOSTORM (Babcock, Sigal, and Zhuang 2012). To discard poorly sampled clusters, localization stitching was used to select clusters with more than 10 ParB-mEos2 molecules. Stitching was realized by grouping together localizations appearing in successive frames or separated by up to 10 frames (to account for fluorescent blinking) and that were 50 nm apart. Stitched localizations were clustered using a voronoi-based cluster identification method (Levet et al. 2015; Cattoni et al. 2017). In total, 990 clusters with an average of $84 \pm 57$ localizations (mean $\pm$ std) were analyzed (Figure S1C). To validate that chemical fixation did not alter the localization of ParB droplets, we measured the position of droplets in small cells as these most often contained two droplets (Figure S1A). ParB-mEos2 droplets observed by PALM displayed the typical localization pattern of ParB-GFP observed in live cells, i.e. a preferential localization near quarter positions along the cell's main axis (Le Gall et al. 2016).

*3D isotropic reconstructions from single-molecule localizations*

Super-resolved images were reconstructed and classified using our Single-Particle Analysis method (Salas et al. 2017) based on iterations of multi-reference alignment and multivariate statistical analysis classification (van Heel, Portugal, and Schatz 2016). N = 990 clusters were sorted into 50 class averages (Figure S1D): 51% (506/990) of the total clusters fell into only eight classes (Figure S1E) which were further analysed for particle reconstructions. To measure the size of the partition complex, we reconstructed the three dimensional structures of the eight most represented classes using the angular reconstitution method (Van Heel 1987) and evaluated their Full-Width at Half Maximum (FWHM, Figure S1F).

*sptPALM*

Single-molecules for sptPALM analysis were localized using MultipleTarget Tracking (MTT) (Sergé et al. 2008). Single ParB-mEos2 localizations were linked to single tracks if they appeared in consecutive frames within a spatial window of 500nm. To account for fluorescent protein blinking or missed events, fluorescence from single ParB-mEos2 molecules were allowed to disappear for a maximum of 3 frames. Short tracks of less than four detections were discarded from subsequent analysis. The mean-squared displacement (MSD) was computed as described in (Uphoff, Sherratt, and Kapanidis 2014). The apparent diffusion coefficient was calculated from the MSD using the following equation: $D = MSD / (4 \, \Delta t)$, where $\Delta t = 9.7 \, msec$.

In all cases, the distribution of apparent diffusion coefficients were satisfactorily fitted by a two-Gaussian function. These fits were used to define the proportions of low- and high-mobility species present for wild-type and mutant strains. Trajectories were then classified as low- or high-mobility using a fixed threshold. The value of the threshold was optimized so that the number of trajectories in each species matched that obtained from the two-Gaussian fit.

Pairwise distances between ParB molecules were computed as described in (Stracy et al. 2015). First, the tracks appearing in a spatio-temporal window of 50 nm and 67 frames were linked together to correct for long-time fluorescence blinking of mEos2 that could lead to biased results towards short distances. Then, the distances between the first localizations of each trajectory were computed. Finally, these distances were normalized by a homogeneous distribution obtained from simulating random distributions of 500 emitters in a 1 μm by 0.5 μm nucleoid space. The threshold to define the shaded region in Fig. 2C (200nm) was obtained from the value in the x-axis where the pair correlation surpassed 1. This value is considerably larger than the size of a ParB condensate measured in fixed cells (Figure 1) because the partition complex moves during acquisition in live sptPALM experiments, leading to an effective confinement region of ~150-200 nm (Sanchez et al. 2015).

## FRAP

ParB-GFP droplets from FRAP experiments were tracked using 3D-DAOSTORM (Babcock, Sigal, and Zhuang 2012). The fluorescence intensity of each ParB-GFP droplet was computed by integrating the intensity of a 318nm x 318nm region (3x3 pixels) centered on the centroid of the focus. The trace in Figure 2E was built by averaging the time-dependent fluorescent decay/increase curves from photo-bleached/unbleached ParB-GFP droplets. Each curve was first normalized by the initial droplet intensity, estimated by the average intensity in the three images prior to photo-bleaching. To estimate and correct for natural photo-bleaching, we used bacteria present on the same fields of view, but whose ParB-GFP droplet had not been photo-bleached.

## Calculation of ParB concentrations

The concentration of ParB molecules per partition complex was calculated as follows. The total number of ParB dimers per cell is estimated at ~850 dimers, with 95% of them residing within a partition complex (Figure 1B). Therefore, there are in average N = 850/3.5*0.95 = 231 dimers per partition complex, given that there are ~3.5 partition complexes per cell in average. The volume of a partition complex can be estimated as V = 4/3*Pi*r³=4.16*10^20 liters, with r being the mean radius of a partition complex (r = 21.5 nm). Thus, the concentration of ParB in a partition complex is estimated as C = N/Na/V = 9.2 mM, with Na being Avogadro's number.

**Bacterial strains and plasmids**

*E. coli* strains and plasmids are listed in Tables S1-S2. Cultures were grown at 37°C with aeration in LB (Miller, 1972) containing thymine (10 µg.ml$^{-1}$) and antibiotics as appropriate: chloramphenicol (10 µg.ml$^{-1}$); kanamycin (50 µg.ml$^{-1}$). For microscopy and stability assays, cultures were grown at 30°C with aeration in MGC (M9 minimal medium supplemented with 0.4 % glucose (glycerol or maltose for microscopy assays), 0.2 % casamino acids, 1 mM MgSO$_4$, 0.1 mM CaCl$_2$, 1 µg.ml$^{-1}$ thiamine, 20 µg.ml$^{-1}$ leucine and 40 µg.ml$^{-1}$ thymine).

DLT3298 (*sspB::kan*) was constructed by P1 transduction from strain JW3197 (Baba et al. 2006). The excision of the FRT-kan-FRT selection cassette from strains DLT3053 and DLT3298 was performed using the pCP20 plasmid (Datsenko and Wanner 2000). The *sspB* gene under the control of the arabinose-inducible $P_{BAD}$ promoter was introduced in DLT3299 by P1 transduction from strain SM997, resulting in DLT3401 (Δ*sspB* $P_{BAD}$-*sspB kan*). The mini-F pCMD6 carrying *parA$_F$-mVenus-ssrA*$_{AEAS}$ was introduced by transformation in DLT3401 to generate DLT3469. The induction was performed with 0.2% arabinose.

The mini-F plasmids expressing the fluorescently tagged Par proteins are derivatives of the wild-type mini-F pDAG114. The double ParB mutant, K191A-R195A, was generated by Quikchange mutagenesis from pJYB212. The plasmid pJYB240 was generated from pJYB212 by exchanging the *mEos2* open reading frame by a PCR-amplified *mTurquoise2* DNA (Goedhart et al. 2012) using the InFusion kit (Clontech). The parB$_F$-*mTurquoise2* gene from pJYB240 was introduced into pJYB243, using *Mfe*I and *Spe*I restriction enzymes, to give

pJYB249. A SsrA tag (SsrA$_{AEAS}$; AANDENYSENYAEAS) to specifically induce the degradation of ParA$_F$ was introduced in frame after the last codon of ParA, generating pCMD6. The ssrA$_{AEAS}$ sequence carries the wild-type co-translational signals between *parA*$_F$ and *parB*$_F$. All plasmid constructs were verified by DNA sequencing (Eurofins).

**Table S1 : Strains list**

(all strains used are derivatives of *E. coli* K12)

| | | | |
|---|---|---|---|
| DLT1215 | *F⁻, thi leu thyA deoB supE rpsL, Δ (ara-leu), zac3051::Tn10* | (Bouet, Bouvier, and Lane 2006) | |
| DLT2887 | DLT1215 / pJYB212 | This work | Figures 1B, 1C, 1D, 2A, 2B, 2C |
| DLT3053 | DLT1215 *Hu-mCherry*, FRT-Kan-FRT | (Le Gall et al. 2016) | |
| DLT3289 | DLT1215 *Hu-mCherry* | This work | |
| DLT3264 | DLT1215 / pJYB216 | This work | Figures 3B, 3C, 3D |
| DLT3298 | DLT3289 *sspB::kan* | This work | |
| DLT3299 | DLT3289 Δ*sspB* | This work | |
| DLT3401 | DLT3299 *nocus-P$_{BAD}$-sspB kan* | This work | |
| DLT3469 | DLT3401 / pCMD6 | This work | Figures 4B, 4C, 4D, 4E, 4F |
| DLT3851 | DLT1215 / pJYB213 | This work | Figures 2D 2E, 2F, 4A |
| DLT3997 | DLT1215/pJYB224 | This work | Figures 3D, S2C |
| JW3197 | BW25113 *sspB::kan* | Keio collection; (Baba et al. 2006) | |
| SM997 | *sspB::cat, nocus-P$_{BAD}$-sspB kan* | Gift from C. Lesterlin | |

**Table S2 : Plasmids list**

| | | |
|---|---|---|
| pCMD6 | pJYB249, *parA$_F$-mVenus-ssrA$_{AEAS}$ parB$_F$-mTurquoise2* | This work |
| pDAG114 | mini-F, *repFIA$^+$, parABS$_F$$^+$, ccdB$^-$, resD$^+$, rsfF$^+$, cat$^+$* | (Lemonnier et al. 2000) |
| pJYB212 | pDAG114, *parB$_F$-mEos2* | (Sanchez et al. 2015) |
| pJYB213 | pDAG114, *parB$_F$-eGfp* | (Sanchez et al. 2015) |
| pJYB216 | pJYB212, Δ*parS* | (Sanchez et al. 2015) |
| pJYB240 | pDAG114, *parB$_F$-mTurquoise2* | This work |
| pJYB243 | pDAG114, *parA$_F$-mVenus* | (Sanchez et al. 2015) |
| pJYB249 | pDAG114, *parA$_F$-mVenus parB$_F$-mTurquoise2* | This work |
| pJYB224 | *parB$_F$_K191A_R195A-mEos2* | This work |

**Key resources table**

| Reagent or resource | Source | Identifier |
|---|---|---|
| Deposited Data | doi:10.17632/c3gymp7vbn.1 | Experimental data types |

# Figures

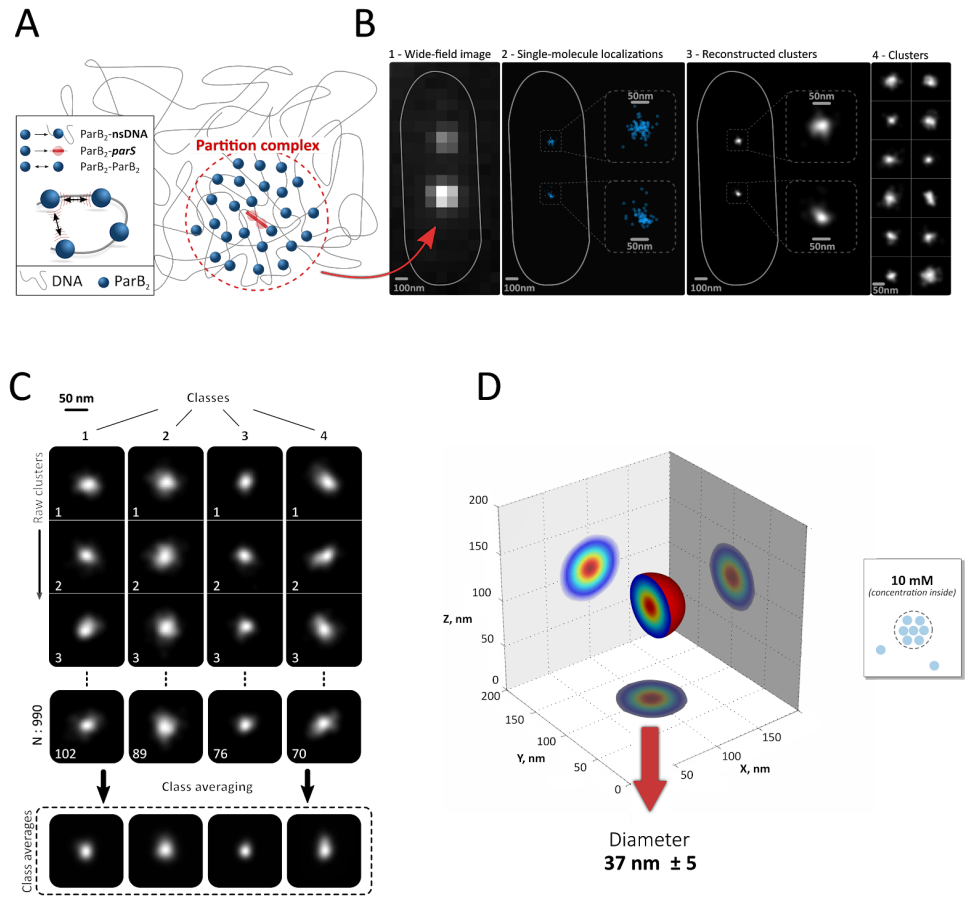

**Figure 1. ParB is confined into spherical nano-condensates. (A)** Schematic representation of a ParB partition complex (red circle). ParB is represented by blue spheres, the centromeric *parS* sequence by a red line, and non-specific DNA (plasmid and bacterial chromosome) is represented by a grey curved line. Three types of interactions have been reported: ParB-*parS* (high affinity, 2 nM), ParB-non-specific DNA (low affinity, 0.3-0.5 μM for DNA fragments of 120-150bp), and ParB-ParB dimer (low affinity, 0.3-1 μM) (Ah-Seng et al. 2009; Sanchez et al. 2015; Taylor et al. 2015). **(B)** Partition complexes (N=990) were visualized by either widefield microscopy (first panel) or by PALM (second and third panels for raw localizations and reconstruction representations, respectively). Magnified views of the two partition complexes are shown for clarity (panel 3). Panel 4 displays representative ParB partition complex reconstructions from other cells to exemplify the heterogeneity in shapes. **(C)** Super-resolved images of ParB partition complexes were classified using single-particle analysis to overcome heterogeneity between single complexes. ParB complexes of the most populated classes exhibited spherical shapes. **(D)** A three-dimensional reconstruction of the ParB partition complex was obtained using the most populated classes. Diameter was estimated by calculating the full width at half maximum. Colors represent reconstruction density (high:red, low:blue), outer shell is represented in shaded red. Right panel: schematic representation of the ParB partition complex and estimation of ParB concentration based on the size of the complex and the number of ParB molecules measured experimentally (see Results and Methods). Measurements were performed in different biological replicates.

Page 34



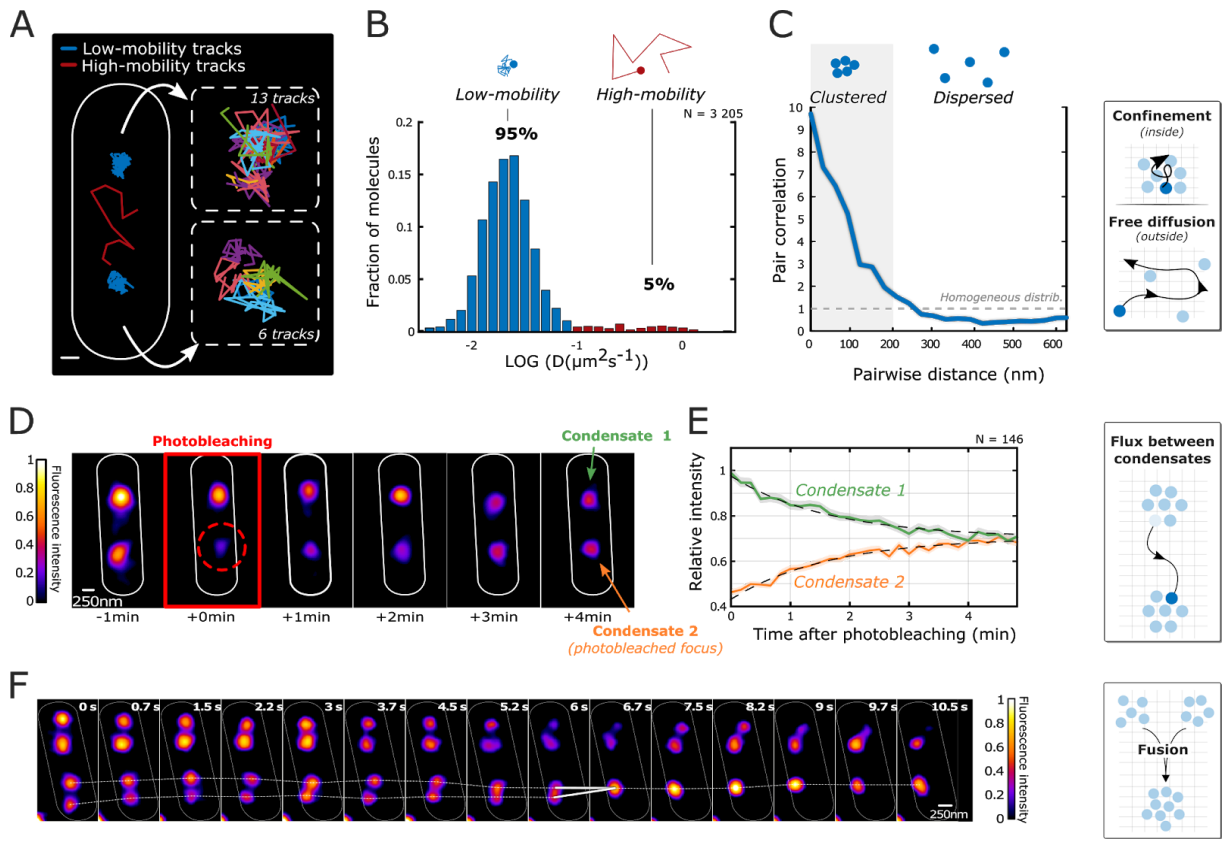

**Figure 2. Dynamic properties of ParB condensates. (A)** Representative image of a live cell (cell contour schematized in white) and trajectories of single ParB molecules (low-mobility: blue; high-mobility: red). Right panels show magnified views of each ParB condensate with different low-mobility trajectories shown with different colors. The slight spatial shifts between trajectories reflect partition complex movement during acquisition (Sanchez et al. 2015). **(B)** Histogram of apparent diffusion coefficients for low- (blue) and high-mobility (red) trajectories. n=156 (number of cells), N = 3205 tracks. **(C)** Pairwise distance analysis of low-mobility trajectories. Expectation for a homogeneous distribution is shown as a horizontal dashed line. Right panels: schematic representations of ParB proteins within (top) and outside (bottom) ParB condensates, the grey matrix represents the DNA substrate. n=156 (number of cells). **(D)** Representative example of a Fluorescence Recovery After Photobleaching experiment. A single ParB-GFP condensate (red circle) was photobleached in a cell containing two condensates. **(E)** The average intensity of photobleached ParB condensates (orange) increased over time with a concomitant and symmetric decrease in the fluorescence intensity of the unbleached condensate (green). This requires single ParB proteins to exchange between liquid and gas-like phases (right panel). Grayed shade represents standard error of the mean. Dashed lines represent the behavior predicted from modeling in Figure S2. n=146 (number of cells). **(F)** Representative example of a fast time-lapse fluorescence movie. Four ParB condensates can be seen to fluctuate and fuse during the movie (white line). Schematic representation of a fusion event is shown in the right panel. Measurements were performed in different biological replicates (n>14 fusion events). In panels A, D and F, solid white lines are used to represent cell contours.





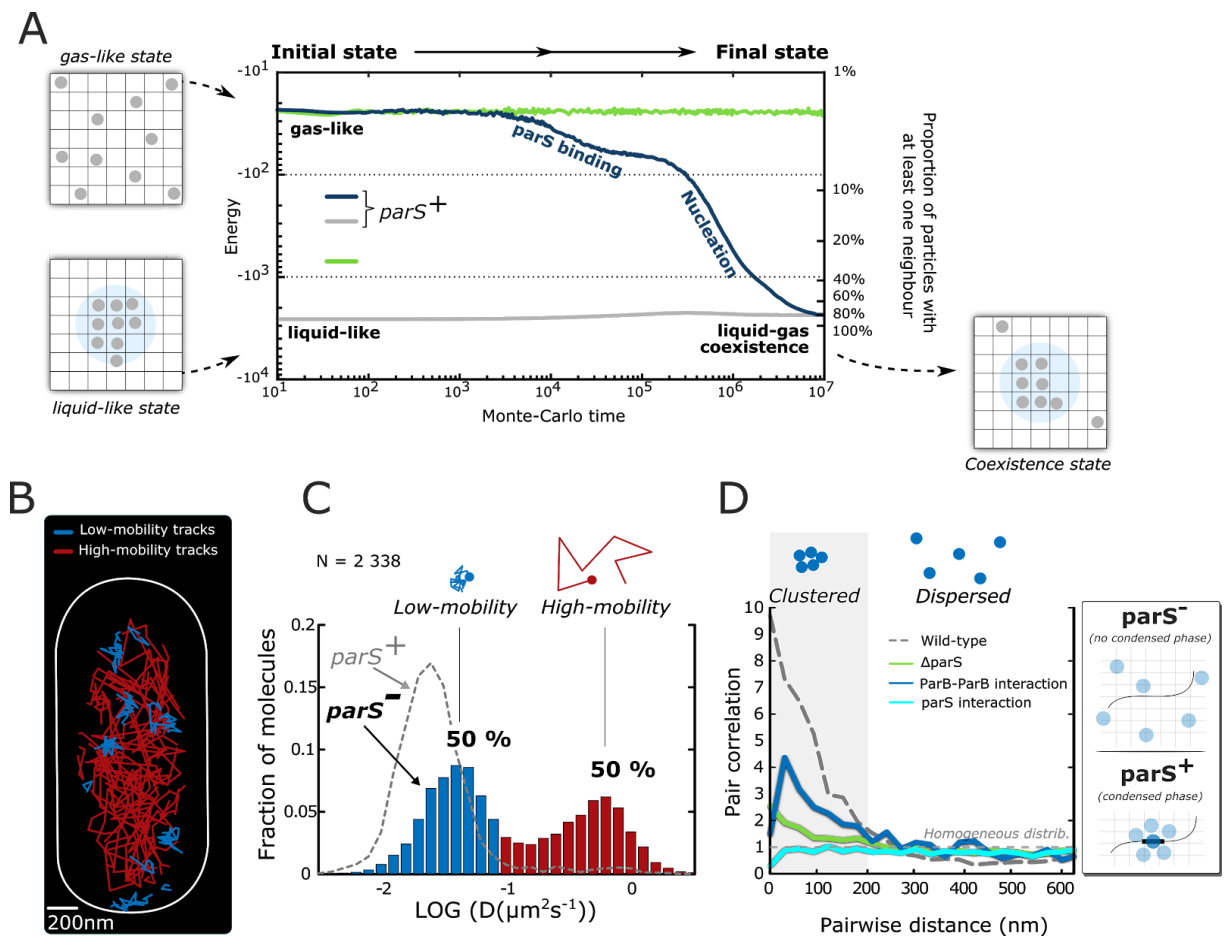

**Figure 3. Centromeric sequences, and ParB-ParB interactions are required for the nucleation and stability of ParB condensates**. **(A)** The movement of three hundred ParB dimers (grey disks) within a simulated cell was governed by free diffusion (1 μm²/s), weak ParB-ParB dimer interactions (J=4.5 kT), high-affinity ParB-*parS* interactions, and low-affinity interactions with the lattice. Regardless of initial conditions, ParB dimers formed a condensate in the presence of *parS10* (gray, blue). In contrast, no condensate is formed in the absence of a nucleation sequence (green). Schematic representations are used to show the initial and final configurations (left and right panels). For simplicity, parS+ refers to a simulation with *parS10*. **(B)** Representative single-molecule tracking experiment of ParB in absence of *parS*. Low-mobility trajectories are shown in blue and high-mobility trajectories in red while the cell contour is schematized as a white line. **(C)** Distribution of apparent diffusion coefficients for low-mobility (blue, 50%) and high-mobility trajectories (red, 50%). The distribution for a wild-type strain containing *parS* (same data as in Figure 2B) is shown as a dashed grey line for comparison. n=209 (number of cells). **(D)** Pairwise distance analysis of low-mobility trajectories in absence of *parS* (green curve), in a ParB mutant impaired for *parS* binding (cyan curve) and in a ParB-ParB interaction mutant (blue curve). The expected curve for a homogeneous distribution is shown as a horizontal dashed line. The distribution for a wild-type strain containing *parS* is shown as a dashed grey line for comparison (same data as in Figure 2C). Schematic representations of single-molecule ParB trajectories in the absence (top) and presence of *parS* (bottom) are shown in the panels on the right. n=209 (number of cells). Measurements were performed in different biological replicates.





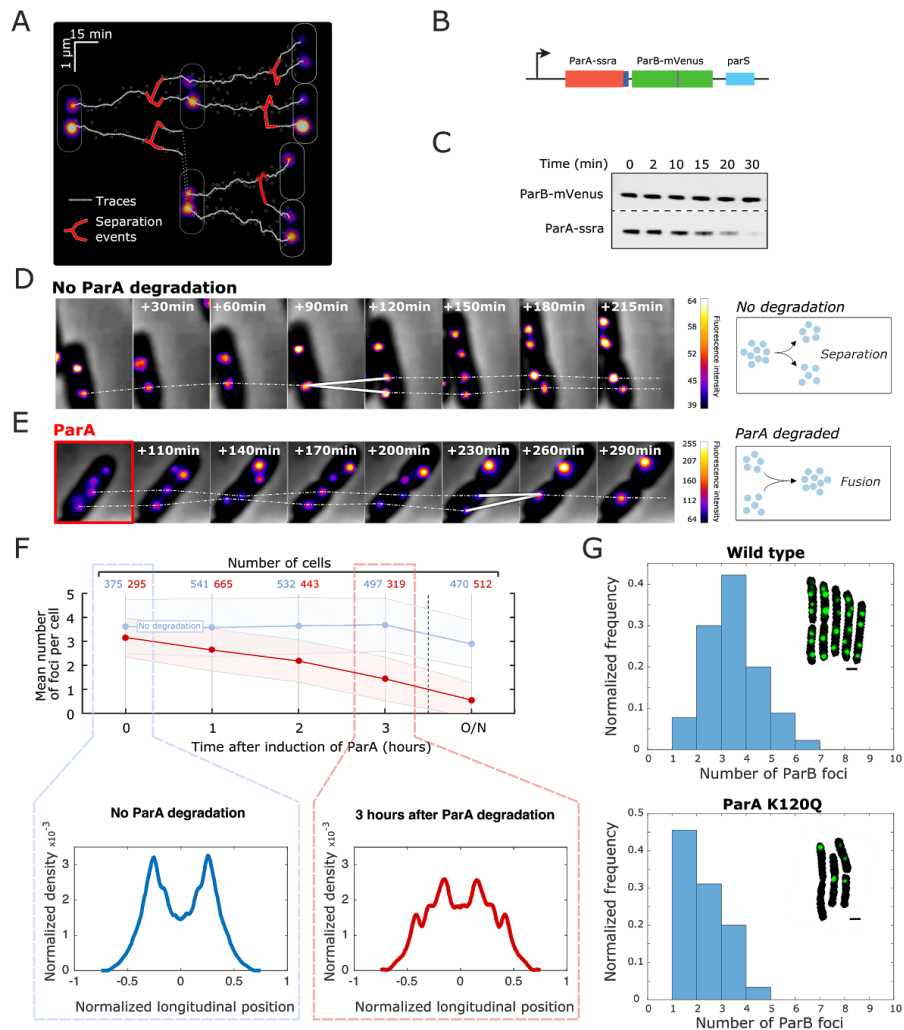

**Figure 4. The ParA ATPase is required for ParB condensates to remain stably separated during cell-cycle times. (A)** Representative kymograph displaying the trajectories of ParB condensates over cell-cycle times (white lines). Splitting of condensates is shown by red forks. The fluorescence images of ParB condensates at three different times as well as the cell schematized outlines are shown for simplicity. **(B)** Genetic construct used for the targeted degradation of ParA. **(C)** Time course Western blot of ParA and ParB, upon induction of ParA-ssra degradation. Degradation time course was detected in two different gels, using anti-ParA and anti-ParB antibodies. **(D)** Representative time lapse experiment of a ParA-ssra strain without induction of ParA degradation. Trajectories of ParB condensates (white dashed line) and a splitting event (white fork) are shown. Scalebar represents fluorescence intensity. Right panel: schematic representation of a splitting event in the presence of ParA. **(E)** Representative time lapse experiment of a ParA-ssrA strain upon induction of ParA-ssra degradation. Dashed lines represent the trajectories of two ParB condensates. A fork displays a fusion event between two ParB condensates. Right panel: schematic representation of this process. Note that time scales in this assay are different from those in Figure 2D (see explanation in Figure S4). **(F)** Quantification of the average number of ParB condensates per cell in a ParA-ssra strain with (red circles) or without (blue circles) ParA degradation. The induction of ParA degradation was initiated at time 0. Shaded areas represent standard deviation. Note that the slight decrease in foci number (~20%) without ParA degradation after overnight (O/N) culture corresponds to the decrease in cell length (~20%) (see Figure S4A). Numbers of cells are shown above plot. Measurements were performed in different biological replicates. Bottom panels: Localization density histograms of ParB condensates along the cell longitudinal axis before (top) and 3 hours after inducing ParA degradation (bottom). **(G)** Distribution in the number of ParB partition complexes in wild-type versus ParA





ATPase deficient cells (ParA_K120Q). The mean number of partition complexes in wild-type cells grown in LB was 3.0±1.1 (N = 105 cells) versus 1.8±0.9 in the ParA ATPase mutant strain (N = 124 cells). Insets show examples of segmented cells (black) with ParB partition complexes (green). Scale bar is 1μm.





# Supplementary Figures

## Figure S1

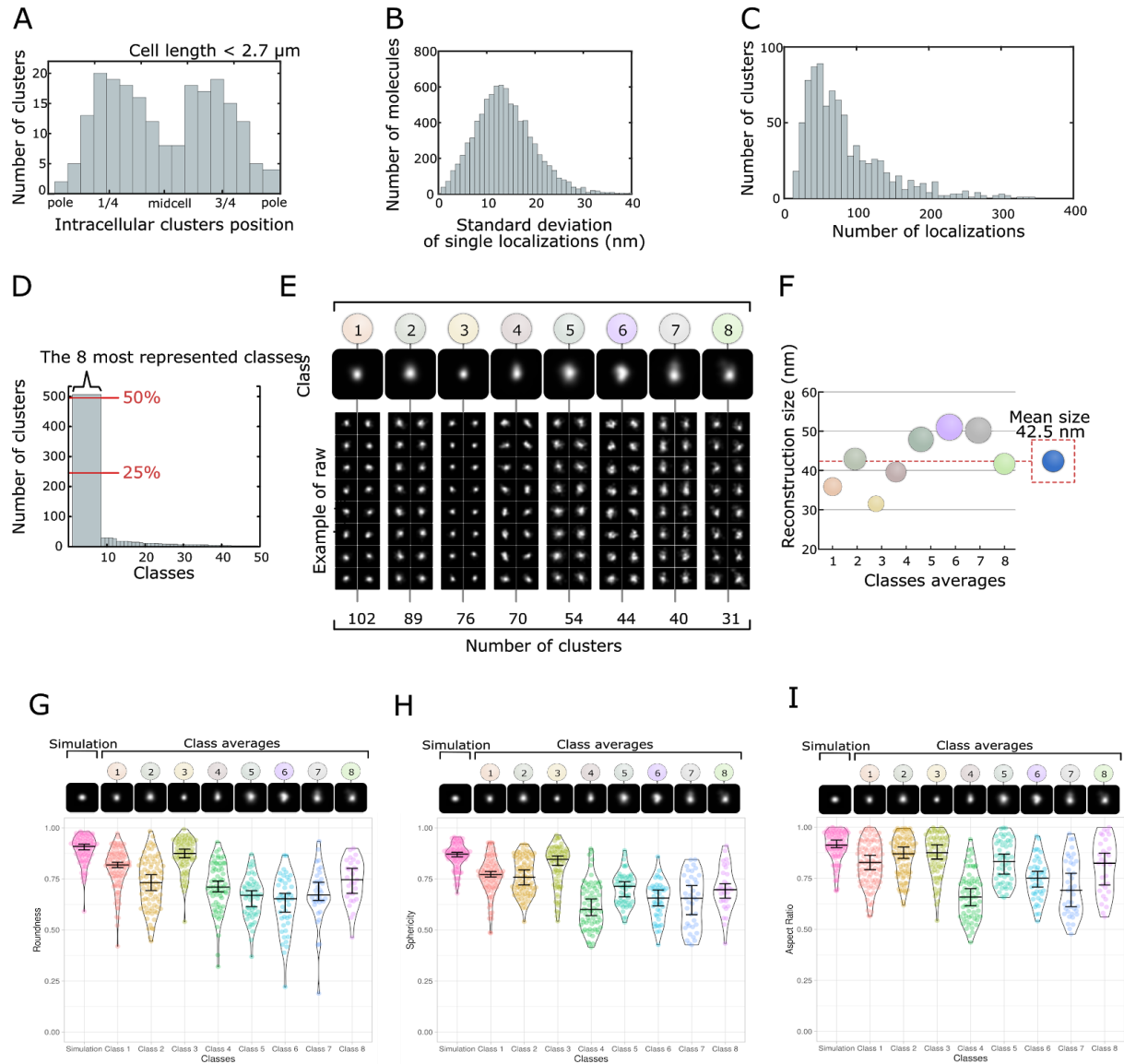

*S1A* To control for fixation artifacts, we built the histogram of localization of ParB clusters along the main axis of the cell. ParB clusters in fixed cells were localized near the quarter positions of the cell axis, mirroring the localization pattern of partition complexes from live experiments (Le Gall et al. 2016). Only clusters of small cells with a length of less than 2.7 μm were selected to avoid cells with more than 2 ParB complexes that would confound the overall localization pattern. N=990 (number of clusters).

*S1B* Histogram of the standard deviations of the localization positions from single molecules. The mean of the standard deviation of localization, which provides an estimate of the localization precision, is $14 \pm 6\ nm$ (mean ± std). N=990 (number of clusters).

*S1C* Histogram of single ParB-mEos2 localizations per cluster. The average number of localizations per cluster is $84 \pm 57$ (mean ± std). N=990 (number of clusters).





**S1E** *Examples of raw clusters for the eight most represented classes, which represent more than 50% of the total number of clusters (N=990).*

**S1F** *Full width at half maxima (reconstruction size) extracted from the 3D reconstructions obtained from each class average. The average reconstruction size is $42.5 \; nm \pm 6.9$ (mean $\pm$ std).*

**S1G-I. Quantitative assessment of particles shape.** *Roundness, sphericity and aspect ratios parameters of individual raw particles were calculated within each class. Particle roundness was computed as the ratio of the average radius of curvature of the corners of the particle to the radius of the maximum inscribed circle, while sphericity was computed as the ratio of the projected area of the particle to the area of the minimum circumscribing circle (Zheng and Hryciw 2015). In addition, we generated simulated 3D localizations homogeneously distributed within a sphere and reconstructed projected density images to obtain a reference of the shape characteristics we should expect for a perfect sphere taking into account experimental localizations errors. The results of these analyses reveal that raw particles display highly symmetric shapes (Panel H) with very little roughness in their contour (Panel G). The residual deviations of some classes from a perfect sphere likely originate from shape fluctuations over time. Finally, we computed the aspect ratio of each particle in each class (Panel I). In this case, raw particles were aligned by using MRA/MSA (van Heel, Portugal, and Schatz 2016), their sizes were estimated in x and y by measuring the full width at half maximum, and the aspect ratio was calculated as a ratio between the particle sizes in the x and y dimensions. This analysis shows that the aspect ratio was on average close to one, as expected for spherical particles.*





## Figure S2

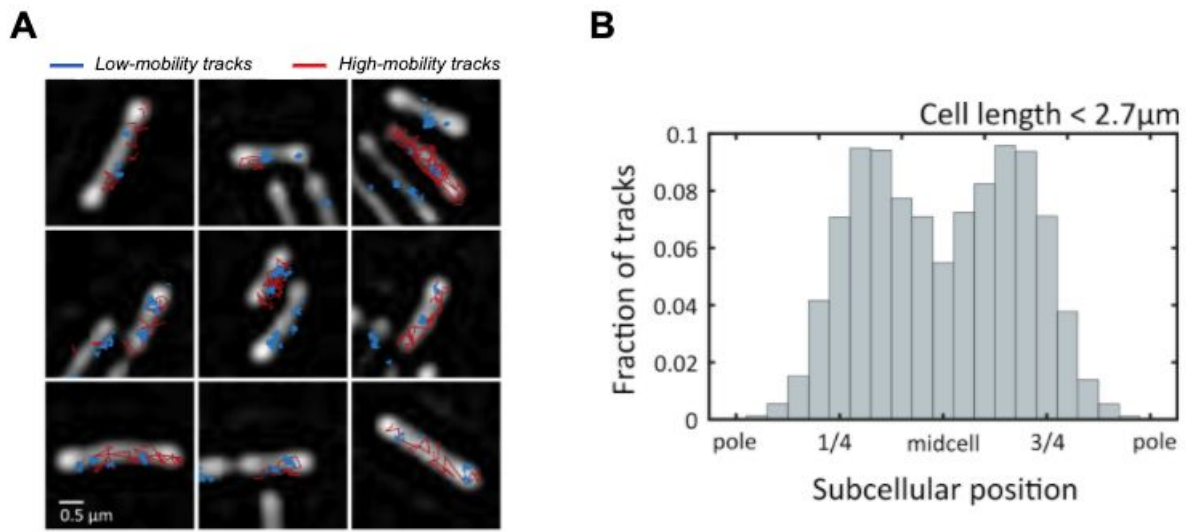

***S2A*** *Representative images of live cells and single ParB trajectories (low-mobility: blue; high-mobility: red). Processed bright field images are shown in white. Bright field images are not drift-corrected, thus in some fields of view trajectories may appear shifted with respect to cell masks. Low-mobility trajectories tend to cluster while high-mobility trajectories display a dispersed nucleoid localization.*

***S2B*** *Sub-cellular localization of ParB tracks along the long axis of the cell.*

<u>Apparent diffusion coefficients for high- and low-mobility species</u>

The apparent diffusion coefficient of proteins bound to DNA non-specifically was reported to range between 0.3 and 0.7 µm2/s (Garza de Leon et al. 2017; Stracy and Kapanidis 2017; Stracy et al. 2016). These values are consistent with the high-mobility species corresponding to ParB particles interacting non-specifically with chromosomal DNA.

The measured apparent diffusion coefficient (D*) comprises cell confinement, motion blurring and localization uncertainty (Stracy et al. 2016). Thus, immobile molecules have a non-zero D* due to the localization uncertainty in each measurement, $\sigma_{loc}$, which manifests itself as a positive offset in the D* value of $\sim \sigma_{loc}^2/\Delta t$ ($\Delta t$ is the exposure time). By using the localization uncertainty and exposure times in our experimental conditions ($\sigma_{loc}$ ~15nm and $\Delta t$ = 8ms), we estimate a non-zero offset in the apparent diffusion coefficient of ~0.02 µm²/s, exactly our estimation of D* for the low-mobility species. Previous studies reported an apparent diffusion coefficient of ~0.1 µm²/s for proteins bound specifically to DNA (Stracy and Kapanidis 2017; Stracy et al. 2016; Zawadzki et al. 2015; Thrall et al. 2017). This D* was similar to that of chromosomal DNA itself (~0.11 µm²/s), as reported by the D* of LacI/TetR specifically bound to DNA (Garza de Leon et al. 2017; Normanno et al. 2015). These studies, however, used different experimental conditions ($\sigma_{loc}$~40 nm and $\Delta t$ = 15ms), leading to a D* for immobile particles of ~0.1 µm²/s (Stracy et al. 2016; Zawadzki et al. 2015; Garza de Leon et al. 2017). Thus, we conclude that the D* of





our low-mobility species is consistent with previous measurements for the apparent diffusion coefficient of proteins bound specifically to their DNA targets.

### Calculation of residence time in ParB condensates

In Figure 2E are shown the fluorescence intensity profiles in time from both ParB condensates after the photo-bleaching of the condensate that we number by 1. We first notice that the foci intensity curves are symmetric (see Figure S2C below) by doing a simple manipulation of the data : $I_1^*(t) = 2I_\infty - I_1(t)$ is the symmetric curve of $I_1$ according to the asymptotic intensity value $I_\infty = I_1(t \to \infty) = I_2(t \to \infty) \approx 0.7$.

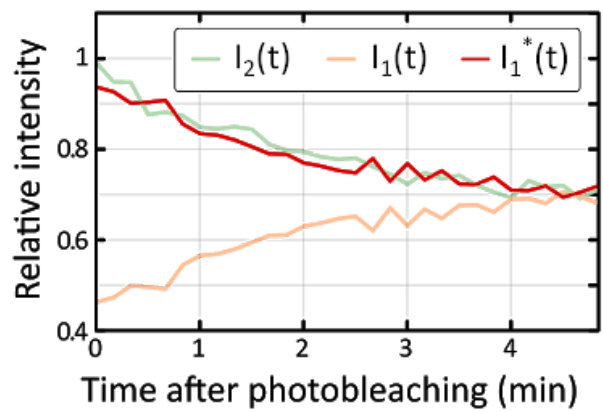

**S2C** Fluorescence intensity curves of photo-bleached (orange, $I_1(t)$) and unbleached condensates (green, $I_2(t)$). The $I_1^*(t)$ curve (red) is symmetric to $I_1(t)$ according to the asymptotic value $I_\infty = I_1(t \to \infty) = I_2(t \to \infty) \approx 0.7$. It shows the symmetrical behavior between both condensates which we used in the model developed here.

We therefore describe the FRAP experiments by the following simple kinetic model on the distribution of ParB proteins in both condensates and the rest of the cytoplasm.

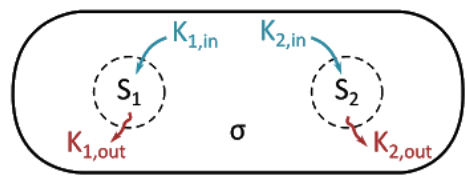

**S2D** Schematic representation of the model parameters. $S_1(t)$ ($S_2(t)$, respectively) is the ratio of the average number of fluorescent ParB-GFP proteins in the photo-bleached (un-bleached, resp.) condensate. $\sigma(t)$ is the ratio of the average number of fluorescent ParB-GFP proteins outside of the condensates. Due to symmetry, we consider : $k_{1,in} = k_{2,in} = k_{in}$ and $k_{1,out} = k_{2,out} = k_{out}$.

This model is described by the following equations





$$\frac{dS_1(t)}{dt} = k_{1,in}\sigma - k_{1,out}S_1(t)$$

$$\frac{d\sigma(t)}{dt} = -(k_{1,in} + k_{2,in})\sigma(t) + k_{1,out}S_1(t) + k_{2,out}S_2(t) \qquad (1)$$

$$\frac{dS_2(t)}{dt} = k_{2,in}\sigma(t) - k_{2,out}S_2(t)$$

where $S_1(t)$ and $S_2(t)$ are, respectively, the ratios of average number of ParB proteins in the condensates (droplets) $F_1$ and $F_2$ after photobleaching and $\sigma(t)$ the ratio of average number of ParB proteins in the cytoplasm after photobleaching. All these quantities were normalized by $S_2(0)$ (see also below) in order to directly compare the experimental normalized intensities $I_1(t)$ and $I_2(t)$ with the model variables $S_1(t)$ and $S_2(t)$. Rates $k_{i,in}$ ( $k_{i,out}$, resp.) ($i = 1, 2$) count for the probability per unit of time to enter (exit, resp.) the condensate $F_i$. Note that here we assume that the fluorescence intensity is proportional to the number of ParB proteins, and that there are no steric/exclusion effects among ParB.

Due to the symmetric behavior of the FRAP signals for each condensate (Figure S2C), we assume that the kinetic rates are the same for each condensate $F_i$. Therefore the equations write:

$$\frac{dS_1(t)}{dt} = k_{in}\sigma(t) - k_{out}S_1(t)$$

$$\frac{d\sigma(t)}{dt} = -2k_{in}\sigma(t) + k_{out}(S_1(t) + S_2(t)) \qquad (2)$$

$$\frac{dS_2(t)}{dt} = k_{in}\sigma(t) - k_{out}S_2(t)$$

Starting from these assumptions, we first study the asymptotic limit of eq. 2 where we notice from Figure S2C that $S_1(t \to \infty) = S_2(t \to \infty) = S_\infty \simeq 0.7$. In this case it is simple to notice that :

$$\frac{S_\infty}{\sigma_\infty} = \frac{k_{in}}{k_{out}} \qquad (3)$$

Live data provided an estimation of this ratio (Figure 2C): $S_\infty/\sigma_\infty = k_{in}/k_{out} \simeq 4.5$.

We then write the explicit time dependence of $S_1(t)$, $S_2(t)$ and $\sigma(t)$. As intensities have been normalized by the natural photo-bleaching, we can first use the conservation of the total number of proteins just after FRAP photobleaching:

$$S_1(0) + S_2(0) + \sigma(0) = S_1(t) + S_2(t) + \sigma(t) = S_{tot}$$

$$(4)$$





It is useful to consider that the same condition is valid all over the acquisition (i.e. we also assume that degradation of ParB is negligible during the 5min-experiment), therefore at long times:

$$S_{1,\infty} + S_{2,\infty} + \sigma_\infty = \left(2 + \frac{k_{out}}{k_{in}}\right) S_\infty = S_{tot}. \tag{5}$$

The conservation of the total particle number (ratio) in eq. 4 allows us to eliminate the equation on the variable $\sigma$ from the equations 2 and write the new set of equations :

$$\frac{dS_1(t)}{dt} = k_{in}S_{tot} - (k_{in} + k_{out})S_1(t) - k_{in}S_2(t)$$

$$\frac{dS_2(t)}{dt} = k_{in}S_{tot} - (k_{in} + k_{out})S_2(t) - k_{in}S_1(t) \tag{6}$$

General solutions of the systems of equations (6) is as follows:

$$S_1(t) = S_\infty - \left[\left(S_\infty - \frac{1}{2}S_+(0)\right)e^{-2k_{in}t} + \frac{1}{2}S_-(0)\right]e^{-k_{out}t}$$

$$S_2(t) = S_\infty - \left[\left(S_\infty - \frac{1}{2}S_+(0)\right)e^{-2k_{in}t} - \frac{1}{2}S_-(0)\right]e^{-k_{out}t} \tag{7}$$

where equations (6) are still symmetrical for the 1 ↔ 2 exchange. To perform a fit of the date, it is useful to consider the symmetric and antisymmetric variables $S_\pm(t) = S_1(t) \pm S_2(t)$ and write :

$$\frac{dS_-(t)}{dt} = -k_{out}S_-(t)$$

$$\frac{dS_+}{dt} = 2k_{in}S_{tot} - (2k_{in} + k_{out})S_+(t) \tag{8}$$

The first equation (8) (in accordance with solutions in eq. 7) gives the simple exponential solution:

$$S_-(t) = S_-(0)e^{-k_{out}t} \tag{9}$$

A simple exponential fit (see Figure S2E below) already provides :

$$k_{out} = (0.0100 \pm 0.0004) \ s^{-1}$$





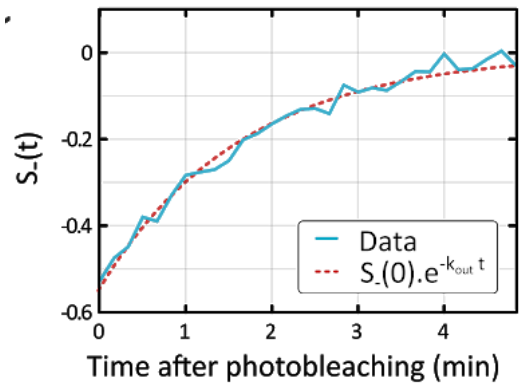

**S2E** *Anti-symmetric intensity curve* $S_-(t) = S_1(t) - S_2(t)$ *(blue) and associated fitted curve* $S_-(t) = S_-(0)\, e^{-k_{out}t}$ *(red).*

or, reciprocally, a residence time in the focus:

$$\tau_{out} = 1/k_{out} \simeq 100s$$

and

$$S_-(0) = (-0.540 \pm 0.022)$$

By using the knowledge of the previously estimated parameters, $k_{out}$ and $S_-(t)$, one can still exploit the solution of $S_2(t)$ (in eq. 7) to fit the remaining parameters $k_{in}$ and $S_+(t)$. We fitted the data with the right side of the following rewritten equation (see fit in Figure S2F):

$$(S_2(t)-S_\infty)e^{k_{out}t} = -\left[\left(S_\infty - \frac{1}{2}S_+(0)\right)e^{-2k_{in}t} - \frac{1}{2}S_-(0)\right] \qquad (10)$$

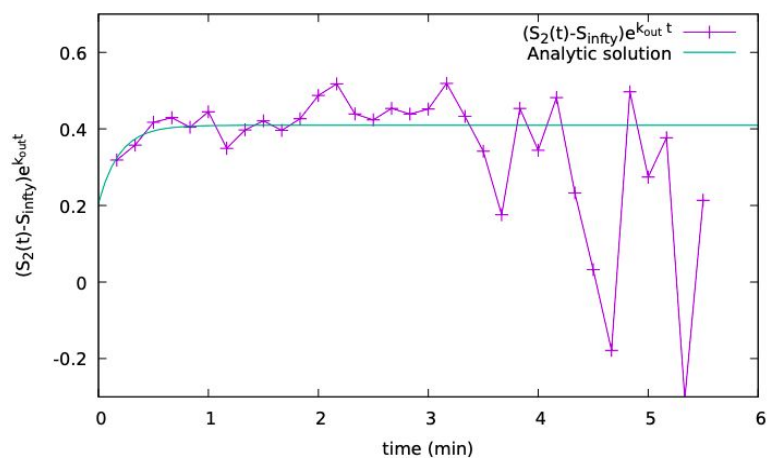

**S2F** *Intensity curve* $(S_2(t) - S_\infty(t))e^{k_{out}t}$ *(purple) and associated fitted curve (see eq. 10)(cyan).*

From this fit, we obtain $k_{in} = (0.044 \pm 0.012)\ s^{-1}$ and $S_+(0) = 0.97 \pm 0.18$. The reciprocal of $k_{in}$ represents the residence time of ParB in the cytoplasm:

$$\tau_{in} = \ \sim 23\ s$$

Remarkably, from this analysis and by using equation 3, we find a ratio of rates:





$$\frac{k_{in}}{k_{out}} = (4.4 \pm 1.3)$$

which provides an estimate (~90%) of the proportion of ParB proteins confined within ParB condensates, in good agreement with experimental results (~95%). The agreement between theoretical and experimental curves can be verified by plotting eqs. (7) together with the experimental datasets (Figure 2F).

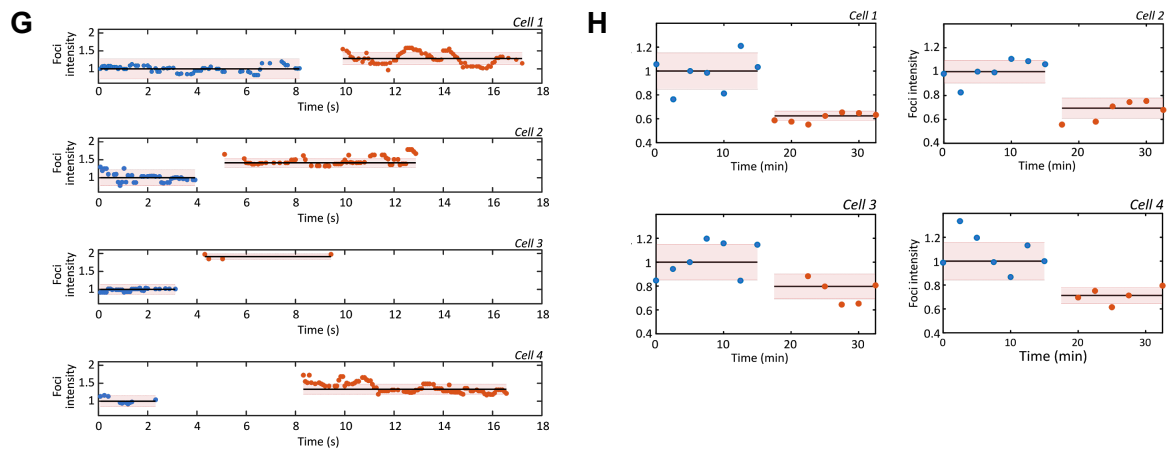

**S2G** *Integrated fluorescence intensity of two ParB condensates before and after fusion. Dots represent the integrated ParB condensate intensity before (blue) and after (red) fusion. Black line and the shaded regions represent median and one standard deviation, respectively. The average increase in fluorescence intensity upon fusion was 1.5±0.1.*

**S2H** *Integrated fluorescence intensity of a ParB condensate before and after splitting. Dots represent the integrated ParB condensate intensity before (blue) and after (red). Black lines and the shaded regions represent median and one standard deviation, respectively. The average decrease in fluorescence intensity upon splitting was 0.7±0.1.*





## Figure S3

The Lattice-Gas model

The ParB proteins in the nucleoid were modeled by a Lattice Gas (LG), which is the paradigmatic system of particles displaying attractive contact interactions (Binney et al. 1992). It is important to note that the LG model presents a first-order phase transition between a gas phase, in which the particles diffuse homogeneously throughout the lattice, and a liquid-gas coexistence phase, in which liquid droplets coexist with diffusing particles in the gas phase. An exchange of particles between the two phases maintains a dynamic equilibrium. As we will see below, the metastable region of the gas can account for the experimental observations on ParBS. We use the LG as a qualitative model for ParB in the nucleoid in order to offer a proof of concept of the mechanism of formation of the ParBS complexes.

For 1d, 2d square, and 3d simple cubic lattices, the LG coordination number is $q = 2d$. Recent work (David et al. 2018) shows that a 1D LG model on a fluctuating polymer like DNA can undergo phase separation and that such a model can, after averaging over the polymer conformational fluctuations, be assimilated approximately to a LG with short range (nearest-neighbor) interactions on a lattice with an effective, perhaps fractal, coordination number between 2 and 6 (G. David, PhD Thesis, in preparation). The increase in coordination number from the value of 2 expected for a linear polymer is due to the combined effects of nearest-neighbor (spreading) interactions along the polymer and bridging interactions between proteins widely separated on the polymer, but close in space. Here for simplicity and purposes of illustration of the basic mechanism we adopt a simple cubic (sc) lattice $(q_{sc} = 6)$.

The ParB proteins were modeled by particles with a diameter $a = 5\ nm$, which were located on the nodes of a simple cubic lattice representing segments of the nucleoid on which ParB is able to bind to or unbind from. The distance between nearest-neighbour sites was as large as the size of a particle, i.e. $5\ nm$. The lattice was chosen to display the dimension of a bacterial nucleoid: $L_x = L_y = 0.5\ \mu m$ and $L_z = 2\ \mu m$ (equivalent to $L_x = L_y = 100\ a$ and $L_z = 400\ a$). The total number of binding sites in the lattice was then $N_s = 4.10^6$. To match the experimentally determined number of ParB proteins per condensate, the particle number was fixed at 300. Thus, the particle density $\rho = N/N_s$ was very low ($\rho \sim 10^{-4}$), placing the LG model in a configuration where only a first-order phase separation transition can occur. Particles could interact between each other when they are at adjacent sites on the sc lattice (nearest neighbour contact interaction) with a magnitude $J$. For the LG model, the total energy $\varepsilon$ of the system is: $\varepsilon = -J \sum_{<i,j>} \phi_i \phi_j$, where $\phi_i$ is the occupation variable at site i taking the value 1 if the site i is occupied or 0 if not. The sum $\Sigma$ runs over the pairs of nearest-neighbour sites <i,j> of the lattice.





### The phase diagram of the LG model

The sc LG model displays only 3 parameters: the particle density $\rho$, the temperature $T$ (or equivalently $\beta$) and the coupling $J$ between particles. Despite its minimalism aspect, the LG model is a paradigmatic model for phase transitions and displays a rich phase diagram (Binney et al. 1992). The metastable region is bound by the coexistence and the spinodal curves, where the LG can transiently remain in a metastable gas or liquid state before crossing a free energy barrier to reach a stable coexistence state. The time required to cross this free energy barrier can be very long, but their transition rate can also be considerably increased by a nucleation factor such as a *parS* site, the effect of which can be modeled within the LG model by particles fixed on the lattice by a strong binding potential. The phenomenology observed in the metastable region is the same as that observed experimentally for a ParB condensate: without a *parS* site, ParBs were homogeneously distributed in the cell, while with *parS* sites, we observed formation of ParB droplets. The *parS* sites catalyzed the condensation of ParB droplets.

Before presenting MC simulations for a finite size system representing the DNA in the bacterial volume, it is instructive to present the thermodynamics of liquid-gas phase separation (infinite volume limit). Although no analytical solution exists for the sc LG with nearest-neighbour interactions, mean field theory provides an approximate analytical solution that gives a global perspective. For a fixed value of $J$, the coexistence and spinodal curves are usually presented in the density-temperature plane ($\rho$, $T$). In the present case it is more useful to fix the temperature at its room value, $T_r$, then plot the coexistence and spinodal curves in the ($\rho$, $J$) plane, with $J$ measured in units of the thermal energy, $k_B T_r$. The mean field phase diagram is presented in Fig. S3A. For the LG model we note the exact (particle-hole) symmetry about the critical density $\rho_c = 1/2$, where the coexistence and spinodal curves coincide.

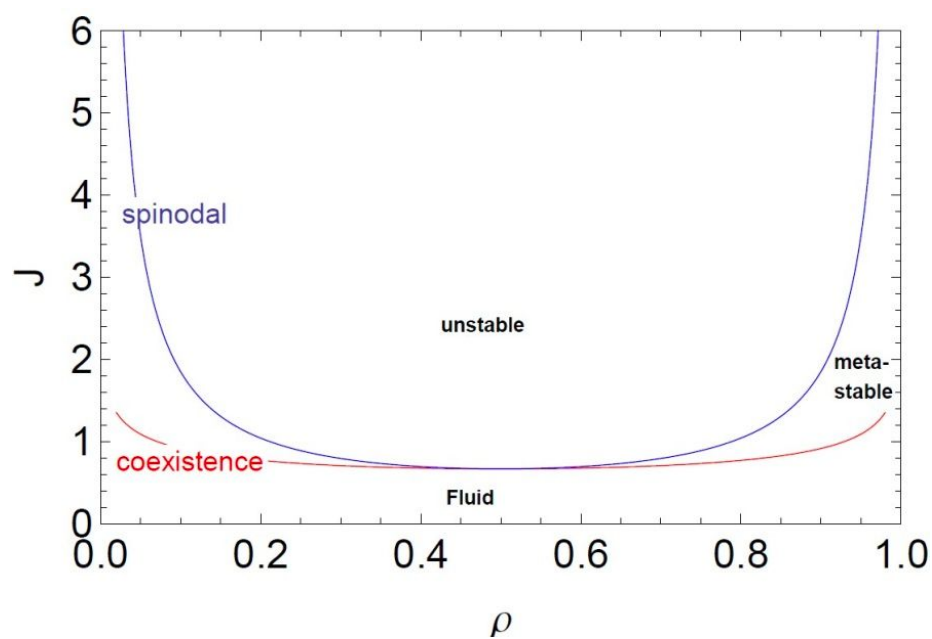





*Fig. S3A. The mean field coexistence and spinodal curves are presented in the density-coupling plane ( $\rho$ , J) with the homogeneous fluid, metastable, and unstable regions indicated.*

At biological density $\rho \sim 10^{-4}$ and room temperature, the homogeneous (possibly meta or unstable) system is in the extreme low-density regime. Although the mean field prediction for the coexistence curve is asymptotically exact in this regime, the same is not true for the mean field spinodal curve. In this limit it is possible, however, to perform an asymptotically exact low density (virial) expansion that leads to a simple expression for the system free energy from which the pressure and chemical potential can be obtained. The coexistence curve is found by equating the pressure and chemical potential in the gas and liquid phases which are in equilibrium. The spinodal curve is determined by the divergence of the isothermal compressibility. At low density $\rho \ll 1$ the gas branch coexistence (coex) and spinodal (sp) curves are given by (G. David, in preparation, 2019):

$$J_{\text{coex}}(\rho) \approx -\frac{2\ln(\rho)}{q_{\text{sc}}}$$

$$J_{\text{sp}}(\rho) \approx -\ln(q_{\text{sc}}\rho)$$

We present in Fig. S3B the asymptotically exact low-density results for the coexistence and spinodal curves in the density-coupling plane ($\rho$, J). The above limiting low density forms indicate that the coexistence and spinodal curves become straight lines in a log-linear plot.

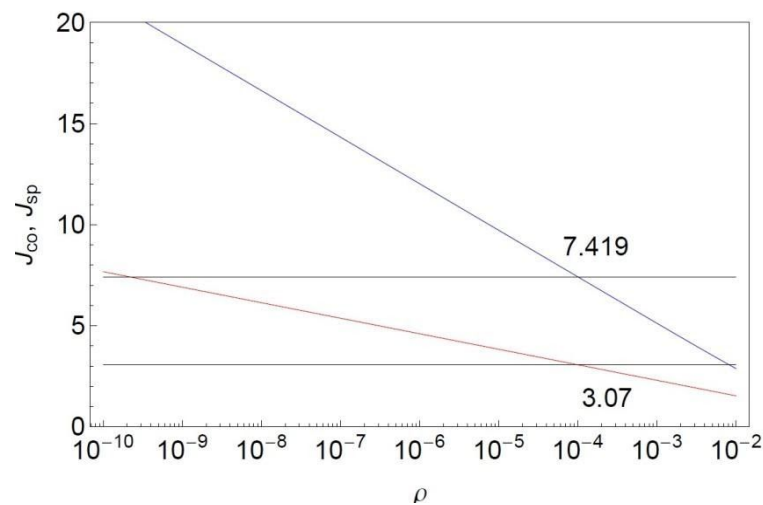

*Fig. S3B. The asymptotically exact low-density results for the coexistence (red) and spinodal (blue) curves in the density-coupling plane ($\rho$, J). The upper and lower limits of metastability in (black horizontal lines) at $\rho = 10^{-4}$ are shown ( $J_{low} = 3.07$ , $J_{up} = 7.42$ ). The homogeneous fluid phase is stable below the coexistence curve, metastable between the coexistence and spinodal curves, and unstable above the spinodal curve.*

A system prepared above the coexistence curve will at equilibrium show phase separation between a low density gas phase with $\rho_{\text{coex}}^{\text{gas}}(J) = e^{-\frac{1}{2}q_{nc}J}$ and a high density liquid phase with (by particle-hole symmetry) $\rho_{\text{coex}}^{\text{liq}}(J) = 1 - \rho_{\text{coex}}^{\text{gas}}(J)$. If the system is prepared in a metastable gas state the time needed to nucleate the phase transition depends on the distance from the coexistence and spinodal curves since the nucleation barrier decreases as the spinodal curve is approached, where it vanishes.

We performed Monte Carlo simulations of the LG in contact with a thermal bath using the standard Metropolis algorithm (Walter and Barkema 2015). In the following, $\beta = 1/(k_bT)$ is the inverse of thermal energy, where T is the absolute temperature and $k_b$ the Boltzmann constant (in the following, all energies are expressed in units of the thermal energy $k_bT$, thus $\beta = 1$). The Monte Carlo procedure corresponding to a unit time step is the following:

1. Choose a particle at random,
2. Propose a move to a random empty site of the lattice.
3. Calculate the energy difference $\Delta\epsilon = \epsilon_f - \epsilon_0$ between the initial and final state,
4. If $\Delta\epsilon < 0$, the displacement is accepted, otherwise the displacement is accepted with a probability $P = e^{-\beta\Delta\epsilon}$,
5. Return N times to step 1.

The particle relocation of the step 2 defines the order of the time unit of the Monte Carlo simulation as approximately one second : ParB proteins diffusing at $D\sim 1\ \mu m^2.s^{-1}$ (Figure 2B) can move at any point on a $1\ \mu m^2$ nucleoid after one second.

We first used simulations to estimate the range of the interaction term $J$ for which the LG model is found in the metastable region, i.e. between the coexistence and the spinodal lines (see Figure S3B and S3C below). To find these lines and define a $J$, we calculated the energy of the system at equilibrium as a function of $J$ with increasing and decreasing variations. Thus, the signature of the first-order transition was visible with a hysteresis cycle and the range of $J$ for which the LG was metastable was bound by the two jumps of energy: the jump at low $J$ corresponding to the coexistence curve and the jump at large $J$ corresponding to the distance from the spinodal curve at which the nucleation time in the presence of a free energy barrier becomes comparable to the simulation time.

To identify the hysteresis cycle, we increased and decreased $J$ by steps of $\Delta J = 0.005$. After incrementing $J$ by $\Delta J$ (or $-\Delta J$ depending on initial conditions), the system was first thermalized to the new coupling value. The sampling of the energy was subsequently performed at intervals long enough to consider that the sampling was not correlated. After $8.10^4$ independent samplings, $J$ was increased to its new value.

We observed different regimes (see Figure S3C below):

I. In the case of a weak initial $J$, the system was first in a gas phase (the interactions are too weak to maintain a cohesive phase). In such a configuration, the system favors a homogeneous distribution of particles to maximize entropy with very few interactions between the particles. Then, by crossing the coexistence line, the gas state became metastable but the system remained in that state until it reached the point before the spinodal line where nucleation time became comparable to the simulation one and the gas phase was depleted in favour of the liquid-gas coexistence phase.

II. In the case of a high initial $J$, the system was first in the liquid-gas coexistence phase: the particles condensed locally in a droplet and increased the number of nearest-neighbour interactions. The droplet coexisted with a small fraction of freely diffusing particles. Then, by approaching and then crossing the coexistence line, the system switched smoothly to a gaseous phase state. This protocol defines a hysteresis cycle determined by simulations for a finite system that delimits the value of J for which the LG is effectively in the metastable phase. Because of the protocol this range depends on the time scale of the simulations. We note furthermore that fluctuations in finite size systems are more important than in an infinite one (thermodynamic limit) and therefore the range of effective metastability determined by simulations should lie within the true thermodynamic range, as is the case.

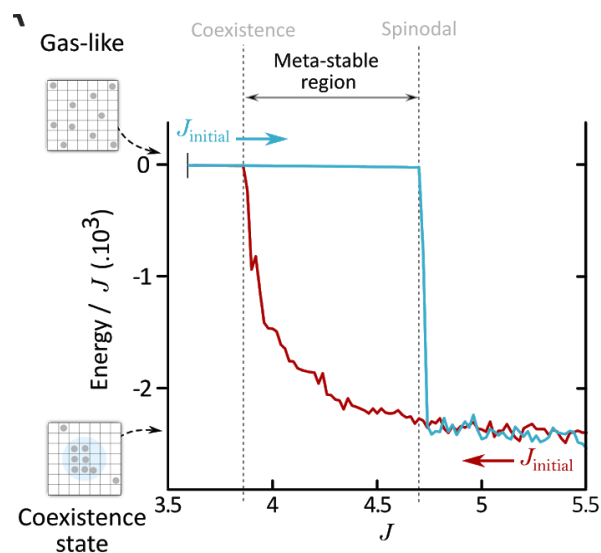

**S3C** *Variation of the energy at the equilibrium of the Lattice-Gas model according to the magnitude of the interaction between particles ( $J$ ), in the presence of parS. For weak $J$ 's, the system is in the gaseous state, while for high $J$ 's, the system is in the liquid state. The transition from one state to another takes place in a metastable region delimited by coexistence and spinodal lines. In this region, the state of the system depends on the initial state, either gaseous (blue curve) or liquid (red curve). We observed a hysteresis cycle which is characteristic of a first-order phase transition of the LG model between the two states. For the simulations shown on Figure 3A, we used $J$ =4.5.*

The coupling range for setting the LG in the metastable regime was estimated as $3.9 < J < 4.7$ . This range falls, as it should, in the thermodynamic range $(3.07 < J < 7.42)$ computed above (see Fig. S3B). In the dynamic simulations of the main text, we chose the value $J = 4.5$ close to the estimated upper limit in order to be as close as possible to the experimental value of 90% of ParB inside the droplets ((Sanchez et al. 2015), and here). This value is in semi-quantitative agreement with other simulation works on ParB-ParB (Broedersz et al. 2014; David et al. 2018). At $J = 4.5$ the gas density on the thermodynamic coexistence curve is very low $\rho_{\text{coex}}^{\text{gas}} \approx 10^{-6}$ (see Fig. S3A) and the liquid density was very close to 1, which leads to 98% of ParB inside the droplets. We expect, however, simulations on finite size systems to show quantitative differences with respect to the thermodynamic limit because of boundary effects and enhanced fluctuations.

<u>Effect of *parS* on the nucleation</u>

The LG provides a proof of concept of the physical mechanisms at work in the formation of ParB droplets: in the metastable region of the liquid-gas coexistence phase, the nucleation factor *parS* site catalyzes the formation of a ParB droplet. To show this effect, we first thermalized the system from a random distribution of ParB on the nucleoid (corresponding to the absence of coupling, i.e. $J = 0$) to the coupling of the simulation $J = 4.5$. At time $t = 0$, we monitored the time evolution of $\varepsilon$ and the number of ParB in the droplet (Figure 3A). The parameters were the same as before, and we averaged over 250 different thermal histories.

We checked two different conditions: **(i)** without a *parS* sequence; **(ii)** with a *parS* sequence, which were modeled by 10 lattice sites with high affinity for ParB proteins (in practice, 10 ParB were always fixed on these sites after a short transitional time). As a control, we also performed a simulation **(iii)** with an initial start where all the particles packed in a single droplet centered on a *parS* sequence. Convergence towards the equilibrium state was achieved in this case by simple relaxation and not by nucleation.

The gas phase is characterized by small values of the energy $\varepsilon \sim 0$ (the particles are dispersed in the lattice) while in the liquid-gas coexistence phase, the condensation of the particles increases the number of interactions, and therefore the energy $\varepsilon << 0$.

In the case **(i)** without *parS*, the system remained at a small equilibrium value $\varepsilon \approx -24 .10^3$ until the final moment of the simulations $t_f = 3.10^7 MC\ steps$, i.e., the condensation into a droplet was not observed. In the case **(ii)**, the system was initially at the small value of $\varepsilon = -24 .10^3$ as in case (i), but was followed by another plateau between $t = 10^4$ and $10^5 MC\ steps$ corresponding to the binding of ParB to the 10 *parS* sites. The energy decreased from $\varepsilon \approx -24$ to $-80 .10^3$, which corresponds to the interactions between the ten ParB fixed side by side on the parS. At the time $t \sim 3.10^5 MC\ steps$, the system switched from a high energy value to a low value, corresponding to the formation of a droplet around *parS*. At larger times $t > 3.10^5 MC\ steps$, the signal saturated at $\varepsilon \sim -2650 .10^3$, indicating that the droplet is stable. Thus, the addition of *parS* reduced the nucleation time by at least two orders of magnitude. These observations correspond to experimental phenomenology: without *parS*, ParB is homogeneously distributed throughout the cell, whereas with *parS*, ParB forms protein condensates.

Case **(iii)** is the control simulation with the fully packed droplet. It took, at large times, the same value of ε as case **(ii)** after a slight decrease corresponding to the escape from the cluster of particles in the gas phase coexisting with the droplet. We also monitored the number of particles inside the droplet (Figure 3A). We observed ~80% of the particles in the droplet, which is in semi-quantitative agreement with experiments (~90% in (Sanchez et al. 2015), and 95% from Figure 2B). In conclusion, the ParB complex behaves similarly to a system in the metastable region of the liquid-gas coexistence phase (Figure S3A). The addition of *parS* acting as a nucleator for ParB condensation significantly reduces the time it takes to form a ParB droplet.

To assess the effect of interaction strengths of ParB particles on their nucleation, we performed additional simulations with lower or higher values of J for our two initial conditions (a gas-like and liquid-like state). When the system was first in a gas phase with weak interaction strengths (J = 3.5 kT, Figure S3D), particles were unable to form condensates regardless of the presence of *parS* sites. Without *parS* sites, particles favored a homogeneous distribution while in presence of *parS* sites particles equilibrated to a state with only ParB particles bound specifically to the centromeric sequence. In the initial conditions where ParB particles were released from a liquid-like state, the system equilibrated to a final state where we did not observe nucleation but rather ParB particles specifically bound to the centromeric sequence.

In the case of strong interaction strengths (J = 6 kT, Figure S3E), with or without *parS* sites, particles initially in a gas-like state immediately started forming small droplets (hence low energy values at initial conditions) until a single large droplet prevailed over the others. In presence of centromeric sequences, no binding of ParB to *parS* was visible due to the fast local aggregation of ParBs at short timescales giving rise to energy variations of the same order of magnitude. In the case where ParB particles were released from a liquid-like state, particles remained in this state throughout the simulation timescales used here. Overall, these simulations highlight the range of interaction strengths required for ParB to nucleate specifically at *parS*. For weak ParB-ParB interactions (low J), ParB particles are unable to form a condensate, regardless of the presence of centromeric sequences. For strong ParB-ParB interactions (high J), ParB particles nucleate spontaneously independently of the presence of *parS* sequences and therefore of the plasmid to segregate. A balance of ParB-ParB interactions strengths and the presence of the centromeric sequence are therefore both essential for the partition complex to form and to function in plasmid partitioning.

The minimalist model presented in this section explains the basic phenomenology of our system, however, it may fail to explain more complex scenarios for which new, more complex models will need to be devised.

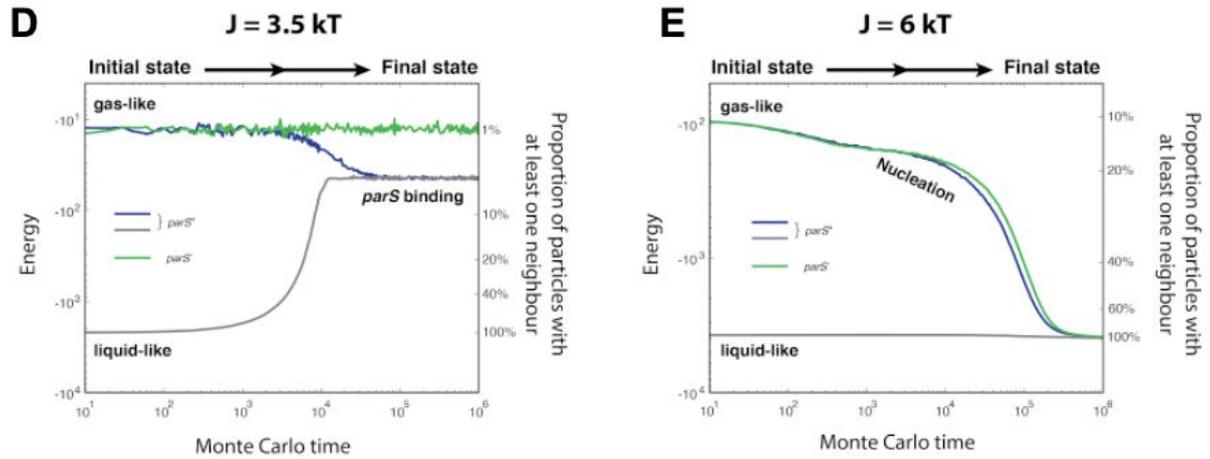

**S3D-E Monte-Carlo simulations of the Lattice-Gas model for weak (left panel) and strong (right panel) interaction strengths between ParB particles. (D)** *Weak ParB-ParB interactions (J=3.5 kT) and high-affinity ParB-parS interactions are considered here. Regardless of initial conditions (gas or liquid-like state), ParB proteins are unable to form a condensate in the presence of parS10 (gray, blue) and the system equilibrates to a state where only ParBs bind specifically to the centromeric sequences.* **(E)** *In contrast, when strong ParB-ParB interactions (J=6 kT) are considered, small droplets immediately appear and nucleate into a single larger condensate independently of the presence or absence of a nucleation sequence (blue and green curves, respectively).*

<u>Nucleation of the partition complex with decreasing number of *parS* sites.</u>

To study the influence of the number of *parS* sites in the speed of nucleation, we performed simulations with a variable number of *parS* sites ($n_{parS}$ = 1 to 10). We observe nucleation for all values of $n_{parS}$, with $n_{parS}$ = 1 and $n_{parS}$ = 2 displaying nucleation times ( $\tau$ ) exceeding the range of MC times reported in the Main Text (see Figure S3F below). However, the nucleation time $\tau$ ($n_{parS}$) follows a power law such that $\tau$ ($n_{parS}$) ~ $1/n_{parS}^2$ (see inset) that can be used to extrapolate the nucleation times for $n_{parS}$ = 1 and $n_{parS}$ = 2. Overall, these data support the finding of Broedersz *et al.* (Broedersz et al. 2014) that a single *parS* site is sufficient to nucleate a ParB condensate, and additionally indicate that the nucleation of ParB condensates is accelerated by the number of *parS* sequences per plasmid.

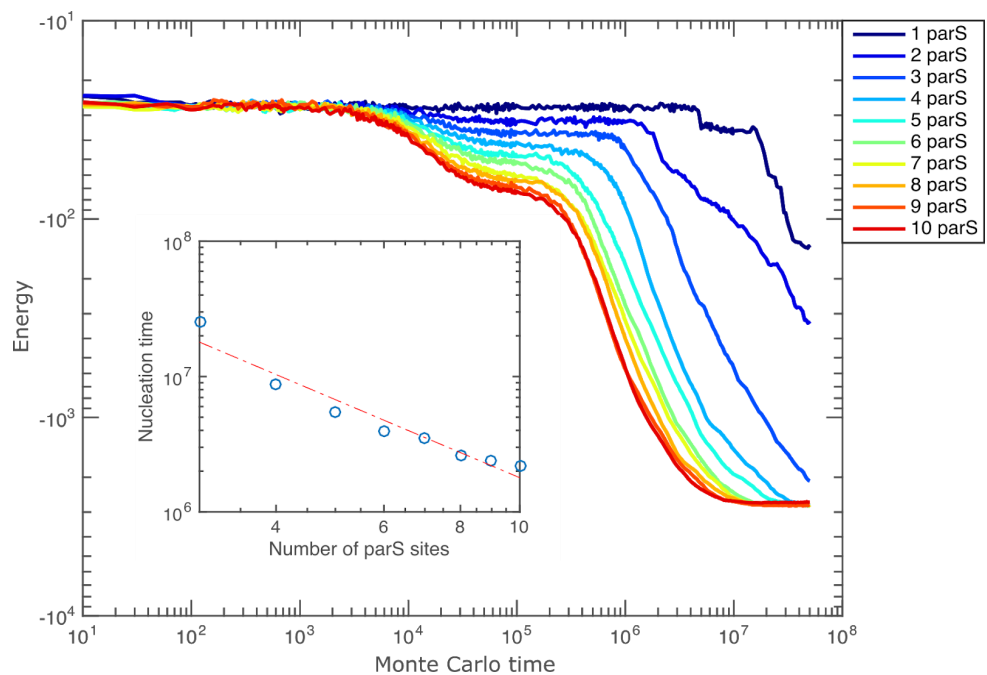

**S3F** *Energy as a function of Monte Carlo time for Lattice-gas simulations using different numbers of parS sites ($n_{parS}$). Nucleation can be detected by a terminal decrease in the energy. Inset: Nucleation time ( $\tau$ ) versus $n_{parS}$. Red dashed line shows a fit to $\tau$ ($n_{parS}$) = $A/n_{parS}^2$, where A is a constant.*

<u>Low-mobility ParB trajectories in the absence of *parS*.</u>

In spt-PALM experiments we observed an unexpected low-mobility ParB population with the strain lacking natural *parS* sequences. The simplest explanation for this result is that, in the absence of *parS*, low-mobility ParB tracks correspond to ParB dimers bound to chromosomal sequences mimicking *parS* (hereby *parS**). This interpretation suggests that in wild-type strains ParB molecules can be in either of three states: (1) in a 'gas-like' state where they move through the nucleoid with low-affinity interactions with non-specific DNA; (2) in a condensed state where they interact specifically with *parS* and with other ParB dimers; (3) in a transient DNA bound state, where they can remain bound to other chromosomal sequences (such as cryptic *parS**) but without interacting with other ParB dimers.

To test this hypothesis, we searched the E.coli chromosome for sequences displaying homology to the *parS* motif (TGGGACCACGGTCCCA, (Pillet et al. 2011)) using FIMO (http://meme-suite.org/tools/fimo).

| Motif ID | Alt ID | Sequence Name | Strand | Start | End | p-value | q-value | Matched Sequence |
|---|---|---|---|---|---|---|---|---|
| 1 | TGGGACCRYGGTCCCA | gi|545778205|gb|U00096.3| | + | 2868055 | 2868070 | 4.27e-07 | 1 | TGTGACCGTGGTCGCA |
| 1 | TGGGACCRYGGTCCCA | gi|545778205|gb|U00096.3| | - | 2868055 | 2868070 | 4.27e-07 | 1 | TGCGACCACGGTCACA |
| 1 | TGGGACCRYGGTCCCA | gi|545778205|gb|U00096.3| | + | 2445771 | 2445786 | 1.39e-06 | 1 | TGGAACCAGGGTCGCA |
| 1 | TGGGACCRYGGTCCCA | gi|545778205|gb|U00096.3| | - | 2445771 | 2445786 | 1.39e-06 | 1 | TGCGACCCTGGTTCCA |
| 1 | TGGGACCRYGGTCCCA | gi|545778205|gb|U00096.3| | + | 590347 | 590362 | 3.4e-06 | 1 | TGGGACCATTGTTTCA |
| 1 | TGGGACCRYGGTCCCA | gi|545778205|gb|U00096.3| | - | 590347 | 590362 | 3.4e-06 | 1 | TGAAACAATGGTCCCA |
| 1 | TGGGACCRYGGTCCCA | gi|545778205|gb|U00096.3| | + | 1857517 | 1857532 | 3.4e-06 | 1 | TCCGACCATGATCCCA |
| 1 | TGGGACCRYGGTCCCA | gi|545778205|gb|U00096.3| | - | 1857517 | 1857532 | 3.4e-06 | 1 | TGGGATCATGGTCGGA |
| 1 | TGGGACCRYGGTCCCA | gi|545778205|gb|U00096.3| | + | 3134505 | 3134520 | 4.95e-06 | 1 | TGGTACGGTGTTCCCA |
| 1 | TGGGACCRYGGTCCCA | gi|545778205|gb|U00096.3| | - | 3134505 | 3134520 | 4.95e-06 | 1 | TGGGAACACCGTACCA |
| 1 | TGGGACCRYGGTCCCA | gi|545778205|gb|U00096.3| | + | 3870876 | 3870891 | 4.95e-06 | 1 | TGGCACCGTGGTCATA |
| 1 | TGGGACCRYGGTCCCA | gi|545778205|gb|U00096.3| | - | 3870876 | 3870891 | 4.95e-06 | 1 | TATGACCACGGTGCCA |
| 1 | TGGGACCRYGGTCCCA | gi|545778205|gb|U00096.3| | + | 1680386 | 1680401 | 6.13e-06 | 1 | TGGGACCATGGGGCGA |
| 1 | TGGGACCRYGGTCCCA | gi|545778205|gb|U00096.3| | - | 1680386 | 1680401 | 6.13e-06 | 1 | TCGCCCCATGGTCCCA |
| 1 | TGGGACCRYGGTCCCA | gi|545778205|gb|U00096.3| | + | 2874741 | 2874756 | 6.93e-06 | 1 | TCGGACCGCTGTACCA |
| 1 | TGGGACCRYGGTCCCA | gi|545778205|gb|U00096.3| | - | 2874741 | 2874756 | 6.93e-06 | 1 | TGGTACAGCGGTCCGA |
| 1 | TGGGACCRYGGTCCCA | gi|545778205|gb|U00096.3| | + | 2881182 | 2881197 | 6.93e-06 | 1 | TGGCAACGCGATCCCA |
| 1 | TGGGACCRYGGTCCCA | gi|545778205|gb|U00096.3| | - | 2881182 | 2881197 | 6.93e-06 | 1 | TGGGACCGCGTTGCCA |
| 1 | TGGGACCRYGGTCCCA | gi|545778205|gb|U00096.3| | + | 2511130 | 2511145 | 9.37e-06 | 1 | TGGGACGGTGGTCACT |

**S3G** *Results of the FIMO analysis for parS mimicking sequences in the E.coli MG1655 genome.*

This analysis shows that there are indeed tens of sequences displaying only 2 or 3 mutations in the consensus *parS* sequence. ParB binds nonspecific DNA with an affinity of 250nM, and *parS* with an affinity of 2.5nM. Introduction of 2-3 mutations in the consensus *parS* sequence changes the affinity of ParB from 6 and 250nM, depending on the position of the nucleotide change (Pillet et al. 2011). Thus, tens of cryptic *parS*\* sequences exist that could be populated when the population of ParB in the gas-like state is abundant.

Next, we performed simulations to explore whether cryptic *parS*\* sequences could get occupied in the absence of the ten natural copies of *parS* in the F-plasmid. For this, we used our Lattice-gas model to measure the occupancy of single copies of *parS* and varied the the energy of interaction between ParB and *parS* (see below).

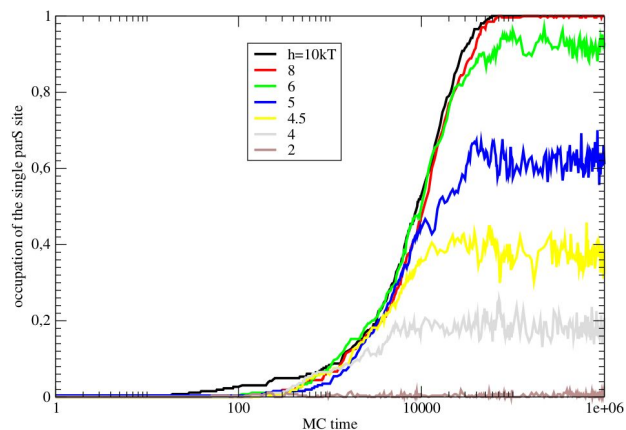

**S3H** *Occupation of single parS sites for different interaction energies as a function of*

*Monte-Carlo time.*

ParB-*parS* interactions with affinities between 6-200nM would have energies between 4.6 and 0.2 kT in our simulations (0 kT representing the baseline for nsDNA interactions). Thus, these simulations show that cryptic *parS** sites would be occupied (0-50%) in absence of the 10 wildtype, high-affinity *parS* sequences, explaining the persistence of low-mobility, unclustered trajectories under these conditions.

Finally, we performed sptPALM experiments on a ParB mutant with reduced *parS*-binding ability to test whether low-mobility trajectories corresponded to ParB molecules bound to cryptic chromosomal *parS** sequences. We observed a reduction in the low-mobility population from 50% in the strain lacking *parS* to 30% in the mutant with reduced *parS* binding ability (Figure S3I). Importantly, residual low-mobility trajectories in this ParB mutant failed to cluster (Figure 3D).

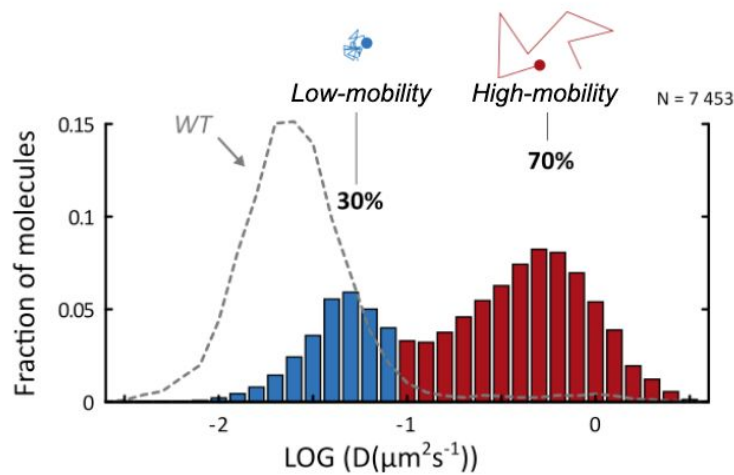

***S3I*** *Distribution of apparent diffusion coefficients for low- (blue, 30%) and high-mobility trajectories (red, 70%) for a ParB mutant impaired in parS binding (pJYB224/DLT3997). The distribution for a wild-type strain (same data as in Figure 2B) is shown as a dashed grey line for comparison. N=7453 tracks.*

This new interpretation explains all our experimental observations: (1) in wild-type cells, ParB dimers in the gas-like fraction can stably bind and nucleate at the F-plasmid because it harbours multiple, high-affinity, contiguous *parS* sequences. This gives rise to low-mobility, clustered tracks observed by spt-PALM. In these conditions, the proportion of ParB dimers in the gas-like state is extremely small, therefore lower-affinity, sparsely distributed *parS** sequences do not get populated; (2) In contrast, in the absence of the 10 wildtype *parS* sequences, a large number of ParB dimers in the gas-like phase can occupy the lower-affinity, sparsely-distributed *parS** chromosomal sites, giving rise to low-mobility, unclustered tracks.

**Figure S4**

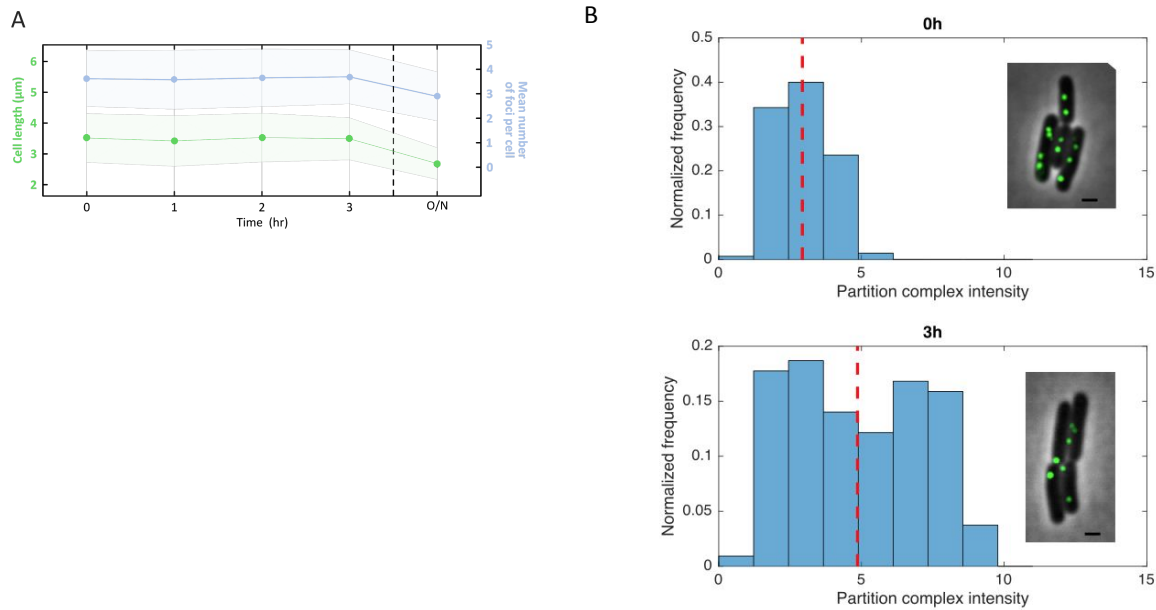

**S4A** *Average cell length and average number of ParB condensates per cell during time-lapse imaging of the ParA-ssrA strain (pCMD6/DLT3469), without induction of ParA degradation. The decrease in the average number of foci per cell observed after one night is due to a change in cellular metabolism, as reflected by the decrease in the average cell length, presumably due to entrance into stationary phase. Shaded areas represent standard deviation.*

**S4B** *Distribution of partition complex mean intensities at t = 0h (top panel, mean intensity: 2.9±0.9) and 3h after ParA degradation in cells where fusion occurred (bottom panel, mean intensity: 4.9±2.3). Insets show a typical field of view. Cells are shown in black and white, and ParB partition complexes in green. Scale bar is 1µm.*

Time-scales for the observation of ParB condensate fusion in Figures 2 and 4

Data presented in Figures 2 and 4 were acquired under very different experimental conditions.

In Figure 2, we used fast time-lapse microscopy (50 ms between frames) to observe fusion of ParB condensates in single cells. From these observations, we found that fusion events occur rapidly (~5±3 s), and perhaps more critically, were in all cases reversible (N>14). With wildtype levels of ParA, the subcellular position of ParB condensates fluctuates rapidly (~few seconds timescale) (Sanchez et al. 2015). Thus, in wildtype cells the random encounter of rapidly fluctuating ParB condensates leads to condensate fusion, and the presence of endogenous levels of ParA ensures their subsequent separation.

In Figure 4, we show that the average number of ParB condensates per cell decreases slowly (over tens of minutes) and irreversibly when ParA is gradually degraded. These measurements were performed every 30 minutes, thus transient and rapid fusion events (as observed in Figure 2) could not be detected due to the sparse and ensemble sampling. In these experiments, ParA degradation is never complete as it competes with ParA expression, therefore, a minimal cellular concentration of ParA may be required to maintain ParB condensates separated at long time-scales.